\documentclass[aps,prb,twocolumn,superscriptaddress,showpacs,10pt]{revtex4-1}
\usepackage{amssymb}
\usepackage{graphicx}
\usepackage{amsmath}
\usepackage{verbatim,dsfont,color}

\begin{document}

\title{Single-spin manipulation in a double quantum dot in the field of a micromagnet}
\author{Stefano Chesi}
\email{stefano.chesi@csrc.ac.cn}
\affiliation{Beijing Computational Science Research Center, Beijing 100084, China}
\affiliation{Center for Emergent Matter Science, RIKEN, Wako, Saitama 351-0198, Japan}

\author{Ying-Dan Wang}
\affiliation{State Key Laboratory of Theoretical Physics, Institute of Theoretical Physics, Chinese Academy of Sciences, P.O. Box 2735, Beijing 100190, China }
\affiliation{Center for Emergent Matter Science, RIKEN, Wako, Saitama 351-0198, Japan}

\author{Jun Yoneda}
\affiliation{Center for Emergent Matter Science, RIKEN, Wako, Saitama 351-0198, Japan}
\affiliation{Department of Applied Physics, University of Tokyo, Bunkyo, Tokyo 113-8656, Japan}

\author{Tomohiro Otsuka}
\affiliation{Center for Emergent Matter Science, RIKEN, Wako, Saitama 351-0198, Japan}
\affiliation{Department of Applied Physics, University of Tokyo, Bunkyo, Tokyo 113-8656, Japan}

\author{Seigo Tarucha}
\affiliation{Center for Emergent Matter Science, RIKEN, Wako, Saitama 351-0198, Japan}
\affiliation{Department of Applied Physics, University of Tokyo, Bunkyo, Tokyo 113-8656, Japan}
\affiliation{Quantum-Phase Electronics Center, University of Tokyo, Bunkyo, Tokyo 113-8656, Japan}
\affiliation{Institute for Nano Quantum Information Electronics, University of Tokyo, Meguro, Tokyo 153-8505, Japan}

\author{Daniel Loss}
\affiliation{Center for Emergent Matter Science, RIKEN, Wako, Saitama 351-0198, Japan}
\affiliation{Department of Physics, University of Basel, Klingelbergstrasse
82, 4056 Basel, Switzerland}

\date{\today}

\begin{abstract}
The manipulation of single spins in double quantum dots by making use of the exchange interaction
and a highly inhomogeneous magnetic field was discussed in [W. A. Coish and D. Loss, Phys. Rev. B {\bf 75}, 161302 (2007)]. 
However, such large inhomogeneity is difficult to achieve through the slanting field of a micromagnet in current designs of lateral double dots. 
Therefore, we examine an analogous spin manipulation scheme directly applicable to realistic GaAs double dot setups. We estimate that typical gate times, realized at the singlet-triplet anticrossing induced by the inhomogeneous micromagnet field, can be a few nanoseconds. We discuss the optimization of initialization, read-out, and single-spin gates through suitable choices of detuning pulses and an improved geometry. We also examine the effect of nuclear dephasing and charge noise. The latter induces fluctuations of both detuning and tunneling amplitude. Our results suggest that this scheme is a promising approach for the realization of fast single-spin operations.
\end{abstract}

\maketitle

% 75.75.-c 	Magnetic properties of nanostructure
% 71.10.Ca 	Electron gas, Fermi gas 
% 75.70.Tj 	Spin-orbit effects 
% 71.23.-k 	Electronic structure of disordered solids

% 71.70.Gm 	Exchange interactions 
% 75.30.Hx 	Magnetic impurity interactions 
% 75.30.Et 	Exchange and superexchange interactions
% 71.10.-w 	Theories and models of many-electron systems
% 73.61.Ey 	III-V semiconductors 

%%%%%%%%%%%%%%%%%%%%%%%%%%%%%%%%

\section{Introduction}\label{sec_general}

Single electron spins confined in quantum dots can constitute building blocks to realize quantum information processing.\cite{Loss1998} The challenges of realizing accurate spin manipulation and the need to achieve easier integration into scalable architectures have stimulated a detailed study of a wide variety of setups and decoherence mechanisms.\cite{Zak2010,Kloeffel2013} In particular, a general strategy to implement a single qubit relies on relatively complex states of several electrons in multiple quantum dots, instead of the spin-1/2 of single electrons. In this approach, it becomes easier to realize single-qubit gates through electric manipulation, at the expense of more cumbersome schemes for the two-qubit gates. A well-studied example is the singlet-triplet (ST) qubit,\cite{Petta2005} based on the spin states $|\uparrow\downarrow\rangle, |\downarrow\uparrow\rangle$ of a double dot, for which universal control and a long lifetime exceeding $200~\mu$s were demonstrated.\cite{Foletti2009,Bluhm2011} Protocols for the CNOT gate were proposed in Refs.~\onlinecite{Taylor2005,Klinovaja2012} and recently an entangling operation of a pair of ST qubits was realized.\cite{Shulman2012}

For the more direct approach of relying on spin-1/2 qubits, the two-qubit operations can be realized on a few-hundred ps time scale\cite{Petta2005} but to achieve selective spin manipulation of individual dots has proved to be more challenging. Electric-dipole spin resonance (EDSR) based on spin-orbit interactions\cite{Golovach2006} was demonstrated with an operation time $\sim 100$ ns in GaAs lateral dots\cite{Nowack2007} and $\sim 10$ ns in InAs nanowire dots.\cite{Nadj-Perge2010} Another promising route relies on the slanting field of a micromagnet,\cite{Tokura2006,Pioro2008} which has allowed coherent rotations with $\sim 100$ ns period\cite{Obata2010} and was integrated with the two-qubit exchange gate.\cite{Brunner2011} Recently, thanks to a better electrical coupling and design of the micromagnet, $\gtrsim 100$~MHz high-fidelity Rabi oscillations were achieved.\cite{Tarucha_micromagnet} However, strong motivations still exist to explore alternative single-spin manipulation schemes, which could achieve a better performance.

An early idea based on inhomogeneous magnetic fields makes use of the spin states  $|\uparrow\uparrow\rangle, |\uparrow\downarrow\rangle$ of a double quantum dot\cite{Coish2007} and can be considered as a compromise between the two strategies outlined above. In fact, the first spin simply acts as an ancillary spin to realize the universal control of the `target' spin through exchange pulses. The two-qubit gates can be realized as usual through the exchange interaction between target spins, with direct tunneling or long-range coupling elements.\cite{Trifunovic2012,Trifunovic2013} The spin manipulation is achieved with pulsed electric control instead of oscillating fields, and $\sim 1$ ns high fidelity gates have been discussed.\cite{Coish2007} 

We consider here a similar approach to control the $|\uparrow\uparrow\rangle, |\uparrow\downarrow\rangle$ states through detuning pulses in a different parameter regime. We specialize to GaAs lateral dots in the slanting filed of a micromagnet but, unfortunately, we find that the conditions of Ref.~\onlinecite{Coish2007} (with negligible hybridization to $|\downarrow\uparrow\rangle$) are not satisfied in current setups. Therefore, we have explored an alternative limit which introduces strong hybridization with the $|\downarrow\uparrow\rangle$ spin configuration and still allows one to achieve $\sim 1$ ns spin rotations in an accessible parameter regime. We have also characterized typical dephasing times within a simple analytic framework. Our analysis suggests that, although strong hybridization of the spin configurations is necessary, the superposition of different charge states and corresponding charge dephasing could be suppressed by a relatively large interdot tunneling amplitude. Such large tunneling also helps to achieve faster rotation times. 

Interestingly, the so-called $S-T_+$ qubit\cite{Ribeiro2010} is based on a quite related spin manipulation scheme. Landau-Zener interferometry through the $S-T_+$ anticrossing, induced by a gradient of the Overhauser field, has been demonstrated.\cite{Petta2010,Ribeiro2013a,Ribeiro2013b} Recent experiments in silicon double dots have explored the same type of Landau-Zener dynamics, but with the anticrossing induced by a micromagnet.\cite{Wu2014} As micromagnets can realize a deterministic slanting field with larger gradient than typical nuclear fields, these advances suggest that single-spin manipulation of the type discussed here offers interesting prospects for a scalable architecture. 

Finally, we note that the use of an additional quantum dot for single-spin manipulation should not be seen as a significant overhead, since the auxiliary dot can be used for efficient readout and single-spin initialization (see, e.g., Sec.~\ref{sec:initializ}).

The paper is organized as follows: In Sec.~\ref{sec:spectrum} we introduce the model and define the logical states. In Sec.~\ref{sec_analysis} we discuss the spin manipulation scheme through an effective two-level system, which clarifies the analytic dependence on system parameters. In Sec.~\ref{sec:numerics} we collect our numerical simulations. In particular, Sec.~\ref{sec:magnet} is on the micromagnet slanting field in a recent experimental geometry, and a simple variation more suitable to our purposes; Sec.~\ref{sec:dynamics} is on the unitary dynamics for the double dot initialization, readout, and spin manipulation; Sec.~\ref{sec:noise} is on the effect of the nuclear and charge noise, which within our approximations can be described with simple analytic expressions. Section~\ref{sec:conclusion} contains our final remarks. Some technical details are presented in the Appendices.

\section{Model and eigenstates}\label{sec:spectrum}

We describe the double dot with the same effective Hamiltonian of Ref.~\onlinecite{Coish2007}. The parameters entering this model can be derived from a more microscopic description following Ref.~\onlinecite{Burkard1999} (see also Ref.~\onlinecite{Stepanenko2012}). $H$ is given by:
\begin{equation}\label{H}
H=H_{C}+H_{T}+H_{Z}.
\end{equation}%
where $H_{C}$ takes into account the electrostatic energies:\cite{vdWiel2002,Coish2007} 
\begin{equation}
H_{C}=-\sum_{l,\sigma }V_{l}n_{l\sigma }+U_{c}\sum_{l}n_{l\uparrow
}n_{l\downarrow }+U_{c}^{\prime }n_1 n_2 .  \label{H_C}
\end{equation}%
$l=1,2$ marks the two dots and $\sigma =\uparrow,\downarrow$ the two
spin directions. The operator $n_{l\sigma }= d_{l\sigma }^{\dag }d_{l\sigma}$ describes the occupation with spin $\sigma$ of the $l$-th dot lowest orbital state, and $n_l=n_{l\uparrow}+n_{l\downarrow }$ gives the total occupation of the $l$-th dot.
The charge configurations are indicated as $(n_1,n_2)$ and we restrict
ourselves to a region in the stability diagram where only (1,1) and (0,2)
are of interest. The first term in Eq.~(\ref{H_C}) is the local
electrostatic potential; the second term is the on-site repulsion, which is
zero for the $(1,1)$ configuration; the third term is the nearest-neighbor
repulsion, which is zero for the $(0,2)$ configuration. As a result, the $%
(1,1)$ configuration has electrostatic energy $E_{\left( 1,1\right)
}=-V_{1}-V_{2}+U_{c}^{\prime }$ and the $(0,2)$ configuration has $E_{\left(
0,2\right) }=-2V_{2}+U_{c}$. As usual, we introduce the detuning parameter $%
\varepsilon =E_{\left( 1,1\right) }-E_{\left( 0,2\right)
}=V_{2}-V_{1}+U_{c}^{\prime }-U_{c}$, which can be controlled through $%
V_l$. If $V_{1}+V_{2}$ is held constant and we set $E_{\left( 1,1\right) }=0$,
then $E_{\left( 0,2\right)}=-\varepsilon $.

$H_{T}$ is the tunneling Hamiltonian between the two dots which is assumed to be
spin-conserving: 
\begin{equation}
H_{T}=t\sum_{\sigma }\left( d_{1\sigma }^{\dag }d_{2\sigma }+d_{2\sigma
}^{\dag }d_{1\sigma }\right),
\end{equation}%
while $H_{Z}$ is the Zeeman energy: 
\begin{equation}
H_{Z}=-\left\vert g\right\vert \mu _{B}\sum_{l=1,2}\mathbf{S}_{l}\cdot%
\mathbf{\ b}_{l},  \label{H_Z}
\end{equation}
with $\mathbf{S}_{l}=\frac{1}{2}\sum_{\rho,\rho^{\prime }} \boldsymbol{\sigma%
}_{\rho,\rho^{\prime }} d_{l\rho }^{\dag }d_{l\rho^{\prime }}$ the spin
operator for the $l$-th dot ($\boldsymbol{\sigma}$ are Pauli matrices) and $\mathbf{\ b}_{l}$ the two local
magnetic fields.  $\mathbf{\ b}_{1}\neq \mathbf{\ b}_{2}$, due to the presence of the slanting field of a micromagnet 
(see Sec.~\ref{sec:magnet}). In Eq.~(\ref{H_Z}) we use a negative $g$-factor, 
appropriate for electrons in GaAs where $g=-0.44$. 

We have neglected in $H$ terms arising from the spin orbit interaction because it is rather weak in GaAs. In fact,
our analysis will yield an energy scale $\Delta E$ for spin manipulation of the order of a few $\mu$eV (e.g., $\Delta E \simeq 3~\mu$eV with parameters as in Fig.~\ref{fig:chargenoise}). On the other hand, the relevant matrix elements of the spin-orbit interaction [modifying Eq.~(\ref{H_matrix})] were discussed in detail in Ref.~\onlinecite{Stepanenko2012}. They are $\lesssim 100$~neV,\cite{Stepanenko2012} thus we expect their effect to be small. Nuclear fluctuations are characterized by a similar energy scale\cite{Stepanenko2012} and indeed we will find in Sec.~\ref{sec:hyperfine} that their effect can be made much smaller than the perturbation due to the micromagnet.

\subsection{Local spin basis}\label{local_spins}

\begin{figure}
\includegraphics[width=0.45\textwidth]{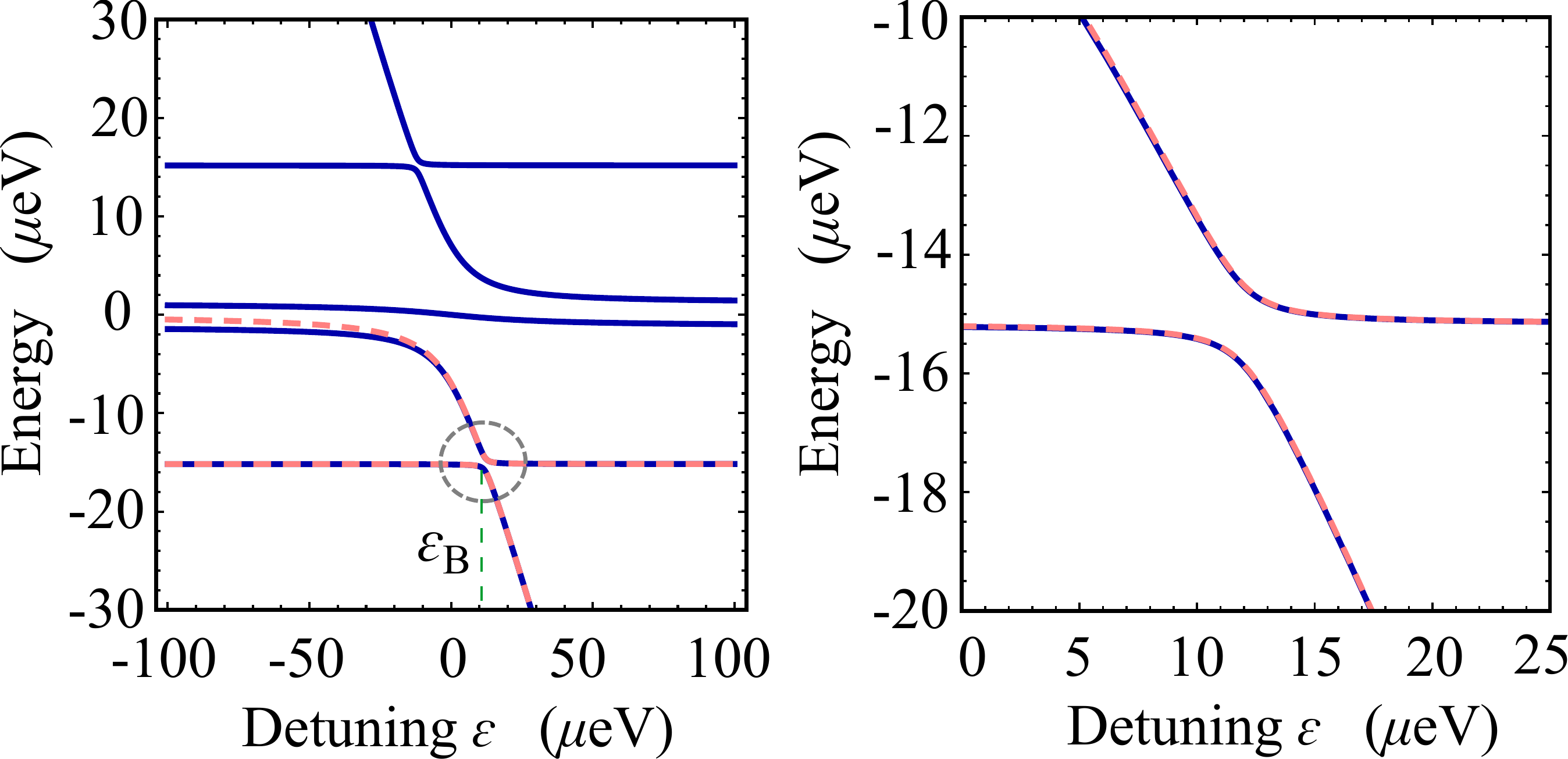}
\caption{Spectrum of the double quantum dot as function of detuning $\varepsilon$. The solid lines are computed from the full Hamiltonian Eq.~(\ref{H_matrix}). 
The dashed circle in the left panel highlights the anticrossing region around $\varepsilon_B$, shown in more detail in the right panel.
The dashed curves are obtained from the effective two-level system of Eq.~(\ref{H_eff}) and describe well the anticrossing region. We used $t=5\,\mu$eV 
and ${\bf b}_{1,2}$ obtained from the geometry of Fig.~\ref{fig:micromagnet}(b) with $B=0.5\,{\rm T}$ and $\varphi=0$.}
\label{fig:spectrum}
\end{figure}

To discuss the properties of $H$, it is useful to introduce the local
spin basis which diagonalizes $H_{C}+H_{Z}$. With a suitable choice
of coordinate axes in the spin space, we can always assume $\mathbf{b}_{1}$ along $z$, while $\mathbf{b}_{2}$ lies in the $x-z$
plane. The angle $\theta $ between the two magnetic fields satisfies: 
\begin{equation}
\cos \theta =\frac{\mathbf{b}_{1}\cdot \mathbf{b}_{2}}{b_{1}b_{2}}.
\label{theta_def}
\end{equation}%
We then define the following operators, where the tilde sign indicates the rotated spin quantization axes: 
\begin{equation}
\left( 
\begin{array}{c}
\tilde{d}_{2+} \\ 
\tilde{d}_{2-}%
\end{array}%
\right) =\left( 
\begin{array}{cc}
\cos (\theta /2) & \sin( \theta /2) \\ 
-\sin(\theta /2) & \cos (\theta /2)%
\end{array}%
\right) \left( 
\begin{array}{c}
d_{2\uparrow } \\ 
d_{2\downarrow }%
\end{array}%
\right) ,
\end{equation}%
while $\tilde{d}_{1+}=d_{1\uparrow }$, $\tilde{d}_{1-}=d_{1\downarrow }$. It is then natural
to discuss the Hamiltonian in the subspace generated by the following basis (with $\mu,\nu=\pm$): 
\begin{eqnarray}
&&|\tilde{\psi}_{\mu ,\nu }\rangle =\tilde{d}_{1\mu }^{\dag }\tilde{d}_{2\nu }^{\dag
}|0\rangle ,  \label{psimunu} \\
&&|\tilde{S}(0,2)\rangle =\tilde{d}_{2+}^{\dag }\tilde{d}_{2-}^{\dag }|0\rangle .
\label{S02}
\end{eqnarray}%
where $|0\rangle$ is the $(0,0)$ charging state. The matrix representation of $H$ is immediately obtained: 
\begin{equation}
H=\left( 
\begin{array}{ccccc}
|g|\mu _{B}b & 0 & 0 & 0 & t\sin \frac{\theta }{2} \\ 
0 & \frac{-|g|\mu _{B}\Delta b}{2} & 0 & 0 & t\cos \frac{\theta }{2} \\ 
0 & 0 & \frac{|g|\mu _{B}\Delta b}{2} & 0 & -t\cos \frac{\theta }{2} \\ 
0 & 0 & 0 & -|g|\mu _{B}b & t\sin \frac{\theta }{2} \\ 
t\sin \frac{\theta }{2} & t\cos \frac{\theta }{2} & -t\cos \frac{\theta }{2}
& t\sin \frac{\theta }{2} & -\varepsilon%
\end{array}%
\right) ,  \label{H_matrix}
\end{equation}%
where $b=(b_{1}+b_{2})/2$, $\Delta b=b_{1}-b_{2}$. $H$ can be diagonalized
easily and an example of the numerical eigenvalues as function of the detuning $%
\varepsilon $ is shown in Fig.~\ref{fig:spectrum} for suitable parameters.

\subsection{Logical states} \label{sec:logical}

The regime of large negative detuning $\varepsilon \ll -t$, is especially
simple since tunneling becomes a small perturbation. More precisely, when 
\begin{equation}
|g \mu_B\Delta b |\gg t^2/|\varepsilon|,
\end{equation}
the effect of $H_Z$ dominates over tunneling and Eqs.~(\ref{psimunu}) and (%
\ref{S02}) become a good approximation of the eigenstates. Similarly to Ref.~%
\onlinecite{Coish2007}, by working at detuning $\varepsilon_A$ deep into
this region, we identify the logical states $|\pm \rangle$ (which we define as eigenstates of the logical $\sigma_z$ operator) 
with the two lowest eigenstates of the double dot. We have:
\begin{equation}\label{factorized}
|\pm \rangle \simeq | \psi_{+,\pm}\rangle,
\end{equation}
where we supposed for definiteness $b_1 > b_2$. In the ideal limit $\varepsilon_A \to -\infty$, $| \pm \rangle$ are products of the spin
states on the two dots and the logical states coincide with the state of the
second spin. An operation in the logical subspace amounts to a single-spin
rotation where the electron in the first dot acts as a frozen ancillary spin.
As the detuning cannot be made arbitrary large, small corrections exist to the
factorized form in Eq.~(\ref{factorized}). These are discussed in Appendix \ref{appendix1}.

\section{Spin manipulation}\label{sec_analysis}

In this section our main result is discussed, i.e., we describe how to realize universal operations 
in the logical subspace and obtain their typical timescales. For a more detailed analysis, including decoherence 
mechanisms, we refer to the following Sec.~\ref{sec:numerics}.

For universal control of $|\pm \rangle$, 
two rotations with independent axes are necessary. The Zeeman splitting at $\varepsilon_A$ provides a natural way
to implement $z$-rotations. By a change of detuning away from $\varepsilon _{A}$ (such that
the system evolves adiabatically in the subspace of the lowest two eigenstates, but the
energy gap between them is modified) a controllable phase shift with respect to the evolution at $\varepsilon_A$ 
can be generated. On the other hand, the $\varepsilon_B$ anticrossing point is of special 
interest, as it allows one to implement rotations about an axis independent of $\hat z$. 
This region, around detuning $\varepsilon_{B}$, is highlighted in Fig.~\ref{fig:spectrum}. 
To characterize the relevant splitting $\Delta E$ at $\varepsilon _{B}$, which determines the rotation timescale, 
we develop a simple analytical treatment of the eigenstates based on first-order
perturbation theory, which is appropriate for the parameter regime of current experiments.

\subsection{Effective Hamiltonian}\label{sec:H_eff}

Our perturbation scheme is based on the fact that in most realizations $\Delta
b\ll b$. In fact, setups involving a micromagnet to generate an
inhomogeneous slanting field\cite{Pioro2008, Obata2010, Brunner2011,
Obata2012} also apply a uniform magnetic field ${\bf B}$ larger
than the saturation field of the micromagnet ($B\gtrsim 0.5$ T\cite%
{Obata2010,Brunner2011}). On the other hand, a field gradient of order $\sim
1$ T/$\mu $m allows one to achieve differences of at most few hundreds mT for
typical quantum dot separations. In practice, the applied $B$ is a few Tesla and
the values of $\Delta b$ realized so far are in the range of $10-80$ mT,\cite{Pioro2008, Obata2010,
Brunner2011, Obata2012, Tarucha_micromagnet} which justifies considering $\Delta b\ll b\simeq B$.
For the same reason, the transverse component (with respect to ${\bf B}$)
of the difference in local fields typically satisfies $\Delta b_\perp\ll b\simeq B$, where:
\begin{equation}
\Delta b_\perp = |(\mathbf{b}_{1}-\mathbf{b}_{2})\times {\bf B}/B|.
\label{b_perp}
\end{equation} 
Thus, $\theta \simeq \Delta b_\perp/b\ll 1$. With the spin coordinates of Sec.~\ref{local_spins} we have in this regime that ${\bf B}$ is approximately along $\hat{z}$ and $\Delta b_\perp \simeq b_{2,x}$.
 
As discussed in more detail in Appendix~\ref{appendix2}, the unperturbed Hamiltonian $H_{0}$ is
simply obtained from Eq.~(\ref{H_matrix}) by setting to zero the diagonal elements proportional to $\Delta b$, as well
as the off-diagonal terms $t\sin(\theta /2)$. $H_0$ can be easily diagonalized
in terms of $|\tilde{T}_{0,\pm}\rangle$, $|\tilde{S}_{\pm}\rangle $ eigenstates, see Appendix~\ref{appendix2}. 
Around $\varepsilon _{B}$ we rewrite $H$ in the $|\tilde{T}_{+}\rangle ,|%
\tilde{S}_{-}\rangle $ subspace, which gives an effective two-level system
described by: 
\begin{equation}
H_{\mathrm{eff}}=\left( 
\begin{array}{cc}
-|g|\mu _{B}b & -\sqrt{\frac{\Delta +\varepsilon }{2\Delta }}\,t\sin (\theta/2) \\ 
-\sqrt{\frac{\Delta +\varepsilon }{2\Delta }}\,t\sin (\theta/2) & 
-\varepsilon /2-\Delta /2%
\end{array}%
\right) ,
\label{H_eff}
\end{equation}%
with
\begin{equation}\label{Delta}
\Delta =\sqrt{\varepsilon ^{2}+8t^{2}\cos ^{2}(\theta /2)}.
\end{equation}%
The detuning at anticrossing is easily obtained: 
\begin{equation}
\varepsilon_{B}=|g|\mu _{B}b-\frac{2t^{2}}{|g|\mu _{B}b}\cos ^{2}(\theta
/2).
\end{equation}%
The eigenstates at $\varepsilon_B$ are 
$|\pm \rangle_{B}=\left(|\tilde{T}_{+}\rangle \pm |\tilde{S}_{-}\rangle \right)/\sqrt{2}$,
with the energy splitting 
\begin{equation}
\Delta E=\frac{2t\sin ( \theta /2)}{\sqrt{1+2\left(\frac{t\cos (\theta /2)}{g\mu
_{B}b}\right)^{2}}}.  \label{DeltaE}
\end{equation}%
As discussed below, $\Delta E$ is the main parameter which deterimins the spin manipulation time. 

\begin{figure}
\includegraphics[width=0.35\textwidth]{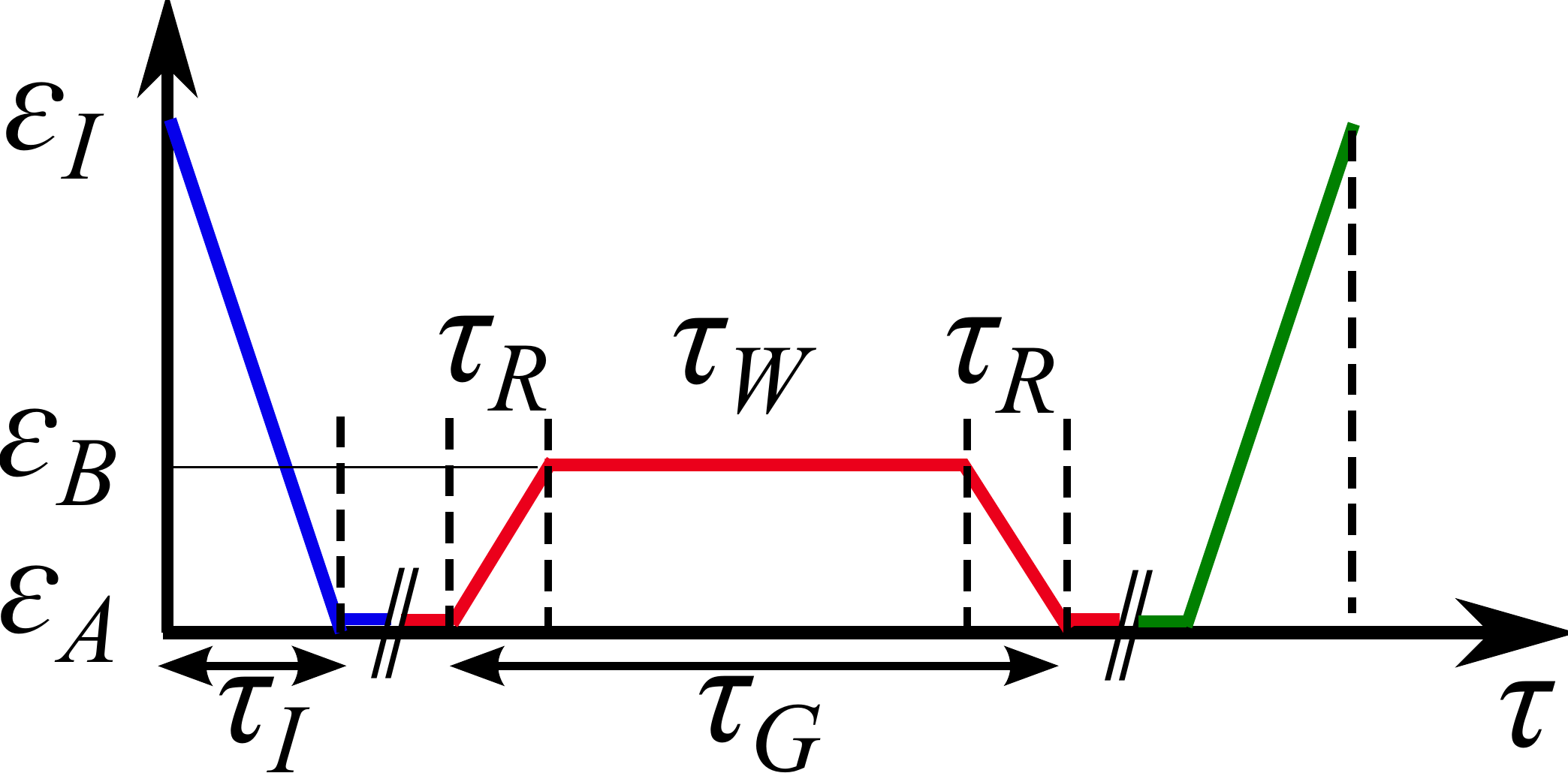}
\caption{Detuning pulses to initialize the system from $\varepsilon_I$ into the logical subspace at $\varepsilon_A$ (first blue pulse), to execute single-spin rotations using the anticrossing at $\varepsilon_B$ (second red pulse), and to read-out the logical states (third green pulse).}
\label{fig:pulses}
\end{figure}

\subsection{Spin rotations}\label{spin_rotation_analytic}
 
We now consider the detuning pulse for spin rotations illustrated in Fig.~\ref{fig:pulses} (in red). The first step is a change in detuning from $\varepsilon_A$ to $\varepsilon_B$, with ramp-time $\tau_R$. Ideally, we would like the evolution to be adiabatic in the lower two energy
branches, except in the vicinity of $\varepsilon_B$. Here, due to the small energy scale $\Delta E$, a diabatic transformation can be realized.
As a consequence, after a time $\tau_R$: 
\begin{equation}
U_R|\pm \rangle \simeq e^{\pm i \phi/2} \frac{|+\rangle _{B}\pm |-\rangle _{B}}{\sqrt{2}},
\label{ideal_sweep}
\end{equation}%
where $\phi$ is a phase which depends on the detailed form of the ramp. 
The $U_R|\pm \rangle$  states evolve for a time $\tau_W$ under $H_{\mathrm{eff}}$ and it is easy to
show that the total effect of the pulse, $U_R^{\dag }e^{-iH_{\mathrm{eff}}\tau_W /\hbar }U_R$, is a
rotation about $\hat{x} \cos\phi+\hat{y} \sin\phi$ by an angle $\Delta E\tau_W /\hbar $. In particular, a $\pi$ rotation is realized when:
\begin{equation}
\tau_W=\tau _{\pi }= \frac{\hbar \pi }{\Delta E},\label{tauPi}
\end{equation}%
which gives $\tau _{\pi }\sim 10$ ns with $b=1$ T,  $t=5\,\mu $eV, and $|g\mu _{B}\Delta b|=1\,\mu $eV 
(we also  assumed $\Delta b_\perp \sim \Delta b$ giving $\theta \simeq\Delta b_\perp/b\simeq 0.04$).

The operation time can be further decreased by increasing the tunneling amplitude, since $\Delta E$ has a significant
dependence on $t$. From Eq.~(\ref{DeltaE}) we have: 
\begin{equation}
\Delta E\simeq \left\{ 
\begin{array}{ll}
2t\sin (\theta /2) & \mathrm{for}~t\ll |g|\mu _{B}b \\ 
\sqrt{2}|g|\mu _{B}b\tan ( \theta /2) & \mathrm{for}~t\gg |g|\mu _{B}b%
\end{array}%
,\right.  \label{DeltaE_approx}
\end{equation}%
which shows that the splitting increases linearly at small $t$, until it
saturates when the tunneling and Zeeman splitting become comparable. The
saturation value of $\Delta E$ gives a lower bound on the operation time. By
taking into account the approximate value of $\theta\simeq \Delta b_\perp/b $:
\begin{equation}
\tau _{\pi }^{\mathrm{min}}\simeq \frac{\sqrt{2}\hbar \pi }{|g|\mu _{B} \Delta b_\perp}.  \label{tauPi_approx}
\end{equation}%
Therefore, the limiting factor for the fastest operation time is given by $\Delta b_\perp$ in this approximation.
Using $|g\mu _{B} \Delta b_\perp|\sim 1~\mu$eV gives $\tau _{\pi }^{\mathrm{min}} \sim 3$~ns. The magnetostatic simulations
of the next section will justify using a larger value for $\Delta b_\perp$, giving $\tau _{\pi }^{\mathrm{min}} \lesssim 1$~ns.
We also notice that, even considering this limit of large $t$ there is typically a clear separation of energy scales $\Delta E\ll |g|\mu _{B}b \lesssim t $ around $%
\varepsilon _{B}$, since $\Delta E\sim |g|\mu _{B} \Delta b_\perp $. Therefore, the pure Hamiltonian dynamics can realize the
operation of Eq.~(\ref{ideal_sweep}) and the associated $\pi$-rotation accurately. A quantitative
characterization of the fidelity for this rotation gate will be discussed later in Sec.~\ref{sec:dynamics} through
numerical simulation.

\section{Numerical studies}\label{sec:numerics}

\begin{figure}
\includegraphics[width=0.48\textwidth]{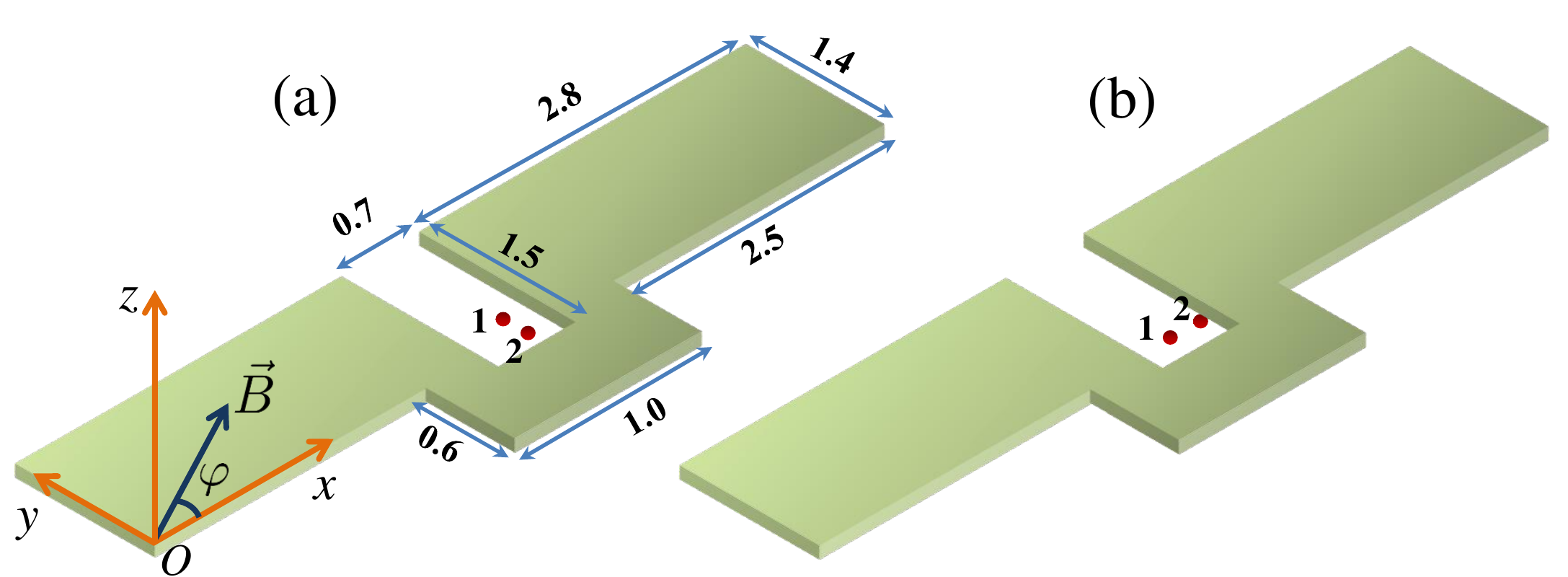}
\caption{Geometries used in the simulations. The micromagnet extends vertically from $z=-0.2\,\mu{\rm m}$ to $z=0$, and is symmetric with respect to the $x=3\,\mu{\rm m}$ plane. Panel (a) indicates relevant dimensions of the micromagnet (in $\mu$m). The coordinates of the two quantum dots are (a) ${\bf x}_1=(3.1,0.4,-0.3)\,\mu{\rm m}$, ${\bf x}_2=(3.1,0.2,-0.3)\,\mu{\rm m}$, and (b) ${\bf x}_1=(3,0.2,-0.3)\,\mu{\rm m}$, ${\bf x}_2=(3.2,0.2,-0.3)\,\mu{\rm m}$. We also indicate the angle $\varphi$ defining the directions of ${\bf B}$ and ${\bf m}$, see Eq.~(\ref{m_direction}). }
\label{fig:micromagnet}
\end{figure}

We now apply the spin manipulation scheme discussed above to a specific setup, which we study by numerical means. In particular, we consider the micromagnet and double dot geometries shown in Fig.~\ref{fig:micromagnet}.
The setup of Fig.~\ref{fig:micromagnet}(a) is very close to the one of Ref.~\onlinecite{Tarucha_micromagnet}, specially designed to optimize the performance of ESR rotations using the micromagnet stray field.\cite{Tokura2006,Pioro2008} We consider simple variations of the geometry and time-dependence of the detuning pulses to represent the typical performance of the spin manipulation scheme. Substantial improvement could be realized by carrying out a more systematic optimization.

\subsection{Slanting field of the micromagnet}\label{sec:magnet}

We consider here magnetostatic simulations\cite{Radia} of the stray field ${\bf b}_{\rm m}({\bf x})$ produced by the micromagnet of Fig.~\ref{fig:micromagnet}, from which the local fields ${\bf b}_{1,2}$ are extracted as follows:
\begin{equation}
{\bf b}_{1,2}= {\bf B} + {\bf b}_{\rm m}({\bf x}_{1,2}),
\end{equation}
where ${\bf x}_{1,2}$ are the centers of the two quantum dots, see Fig.~\ref{fig:micromagnet}. The values of ${\bf b}_{1,2}$ allow us to estimate through Eq.~(\ref{tauPi}) the typical timescale for spin manipulation (see further below). We will also use the obtained values of ${\bf b}_{1,2}$ for our simulations of the spin dynamics in the following subsections.

To obtain ${\bf b}_{\rm m}({\bf x})$, we assume uniform magnetization appropriate for Cobalt, $\mu_0\mathbf{|m|}=1.8\,{\rm T}$ (with $\mu _{0}$ the vacuum permeability). The magnetization direction is determined by the external field ${\bf B}$ and we simply assume that ${\bf m}$ is parallel to ${\bf B}$. These approximations are justified if the magnetic field is sufficiently strong, such that the micromagnet is fully magnetized  and shape anisotropy effects can be neglected. For simplicity, we restrict ourselves to a magnetic field in the two-dimensional plane of the lateral quantum dots, where $\varphi$ is the angle with the $x$-axes (with coordinates as in Fig.~\ref{fig:micromagnet}), thus: 
\begin{equation}
\mathbf{m}=|{\bf m}|(\hat{x}\cos \varphi +\hat{y}\sin \varphi ).
\label{m_direction}
\end{equation}

As the original design of Fig.~\ref{fig:micromagnet}(a) was intended for ESR manipulation based on a resonant electric drive displacing the dots along $x$, the field derivatives in this direction are rather large (we obtain $\partial b_{{\rm m},z}/\partial x \simeq 1.5\,{\rm mT/nm}$ at the position of dot $1$). To take better advantage of the micromagnet geometry, we consider a simple variation in which the two quantum dots are aligned in the $x$-direction instead of along $y$, see Fig.~\ref{fig:micromagnet}(b). In this configuration faster spin rotations can be implemented with the alternative scheme discussed here.

\begin{figure}
\includegraphics[width=0.45\textwidth]{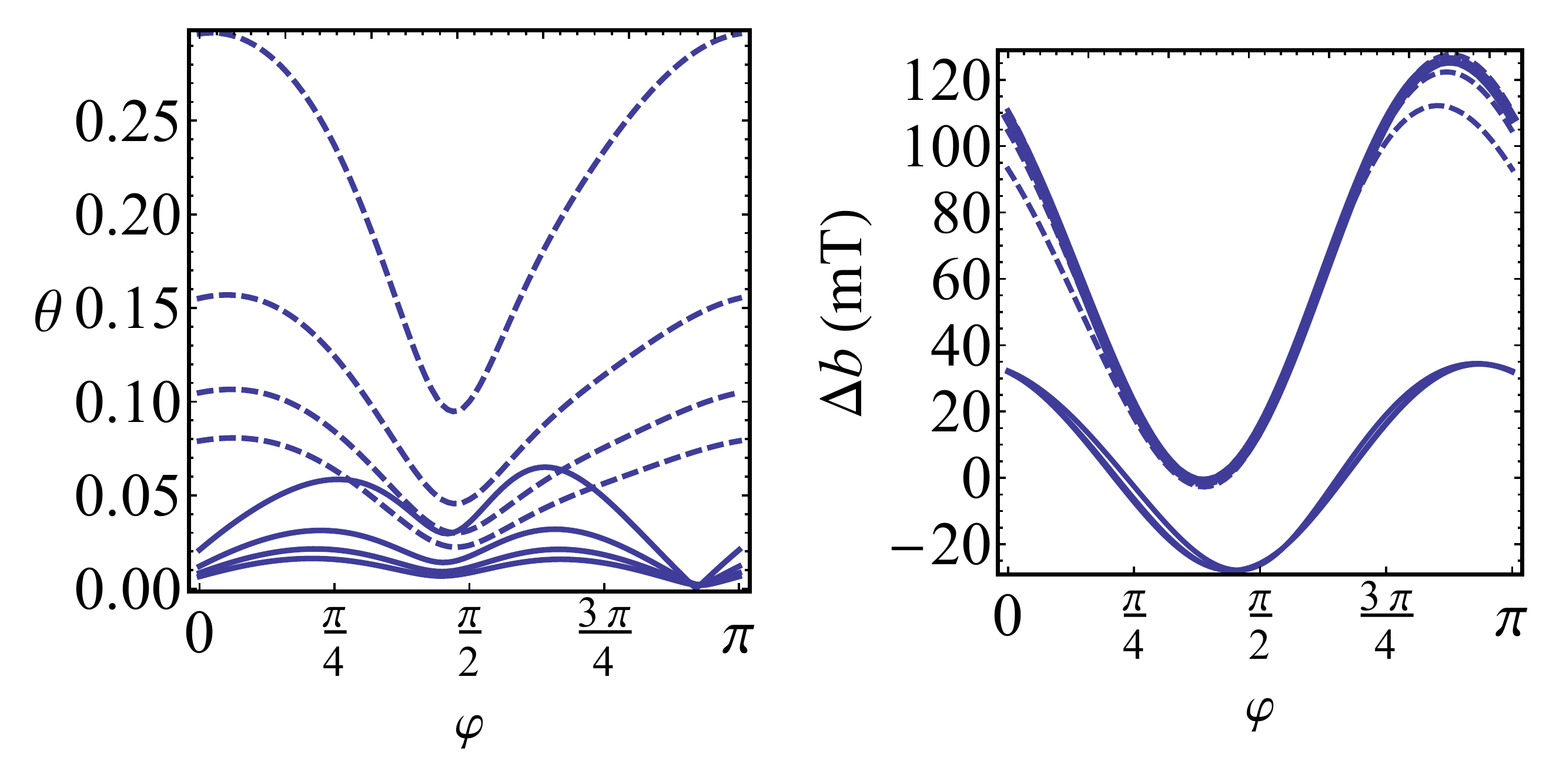}
\caption{Difference in direction $\theta$, see Eq.~(\ref{theta_def}), and magnitudes, $\Delta b=b_1-b_2$, between the local magnetic fields at the two quantum dot locations, as functions of the direction of ${\bf B}$. Solid (dashed) curves are for the setup of Fig.~\ref{fig:micromagnet}(a) (Fig.~\ref{fig:micromagnet}(b)). Each family of curves (solid/dashed in the left/right panel) are for $B=0.5,1,1.5,2$ T from top to bottom, except the dashed curves of the right panel (where $B=0.5,1,1.5,2$ T from bottom to top). }
\label{angle12_simulation}
\end{figure}

In particular, the angle  $\theta$  of Eq.~(\ref{theta_def}) is an important parameter since the anticrossing gap is $\Delta E \sim t \theta$ at small tunneling, see Eq.~(\ref{DeltaE_approx}). The value of $\theta$ is plotted in the left panel of Fig.~\ref{angle12_simulation} as a function of the magnetization direction and with several values of the external magnetic field $B$. As expected, the largest values of $\theta$ are obtained at the smaller external field $B=0.5 \, {\rm T}$. Furthermore, there is a marked dependence on $\varphi$, with an optimal value around $\varphi \simeq 5\pi/8$ for setup (a) and at $\varphi \simeq 0$ for setup (b). We will usually choose these two optimal values in the following, when discussing setups (a) and (b). Furthermore, it is also clear from Fig.~\ref{angle12_simulation} that the values of $\theta$ obtained from setup (b) are significantly larger than those from setup (a).

The advantage of setup (b) in realizing faster rotations can also be seen by computing $\tau_\pi$ with $\theta,b$ obtained in the simulation (we use $b\simeq B=0.5$~T). For setup (a) with $\varphi=5\pi/8$ we obtain $\tau_\pi \simeq 4\,{\rm ns}$ for $t=10\,\mu{\rm eV}$ and $\tau^{\rm min}_\pi \simeq 2.5\,{\rm ns} $ for $t\gg 10\,\mu{\rm eV}$ [see Eqs.~(\ref{tauPi}) and (\ref{tauPi_approx}), respectively]. Therefore, the original design (a) already yields relatively fast timescales for single-spin manipulation. By considering setup (b) with $\varphi=0$ and $t=10\,\mu{\rm eV}$ we obtain an improved operation time of $\tau_\pi \simeq 1\,{\rm ns}$.

Finally, another advantage of the modified setup (b) is the larger value of $\Delta b$ which, as discussed in Appendix~\ref{appendix1}, helps to disentangle the two spins at large negative detuning, thus to realize more faithful single-spin rotations. The values of $\Delta b$ are plotted in the right panel of Fig.~\ref{angle12_simulation} for the two geometries. Similarly to $\theta$, $\Delta b$ has a strong variation with $\varphi$. The dependence on $B$ is much less pronounced as it is mainly determined by the fixed difference in the longitudinal components (i.e., parallel to ${\bf B}$) of the micromagnet slanting fields at the two dot locations.

\subsection{Unitary dynamics}\label{sec:dynamics}

We now consider the unitary dynamics under the detuning pulses illustrated in Fig.~\ref{fig:pulses}. The effect of the anticrossing at $\varepsilon_B$ on initialization and readout is discussed. We also confirm that fast single spin manipulation with gate time $\sim 1$~ns could be realized with high fidelity. The influence of decoherence mechanisms is discussed in the following Sec.~\ref{sec:noise}.

\begin{figure}
\includegraphics[width=0.4\textwidth]{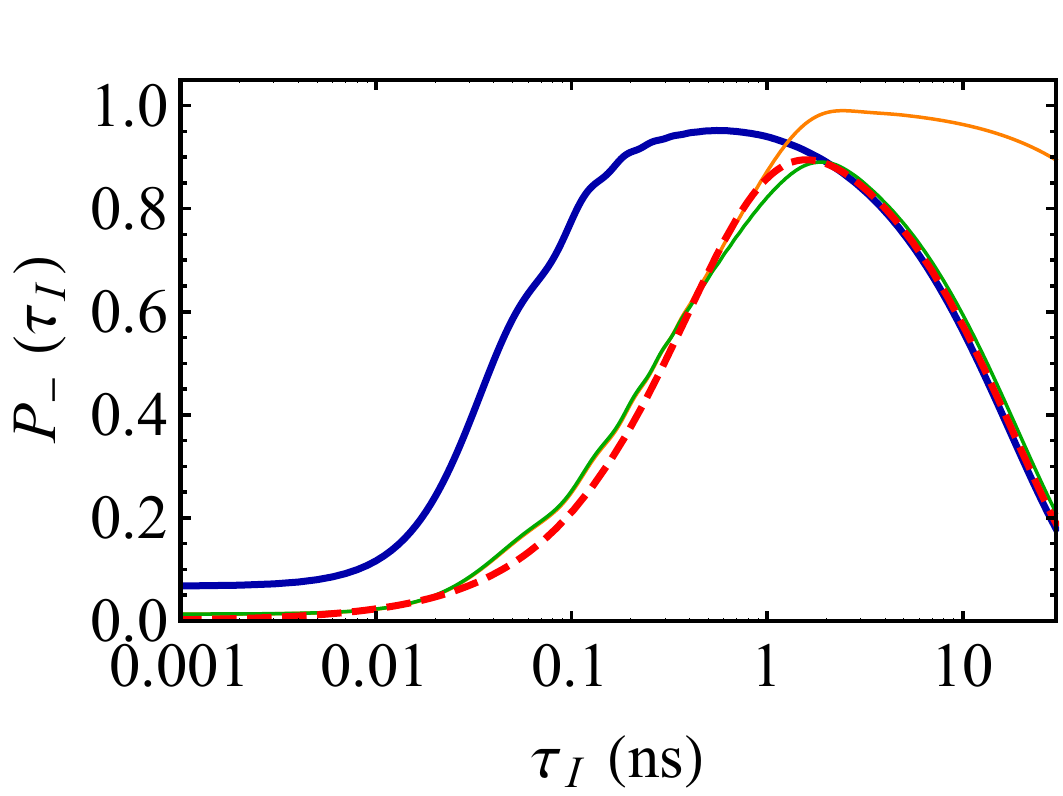}
\caption{Plot of the initialization fidelity $P_-(\tau_I)=|\langle - |\psi(\tau_I)\rangle|^2$ for a linear ramp in detuning $\varepsilon(\tau)$ starting at $ \varepsilon_I = -\varepsilon_A= 200\, \mu$eV (with $|\psi(0)\rangle$ the ground state) and ending at $\varepsilon(\tau_I)=\varepsilon_A$. We used the geometry of Fig.~\ref{fig:micromagnet}(b) with $\varphi=0$. The thin solid curves are for $t=10\,\mu$eV where the lower green (upper orange) curve is for $B=0.5$~T  ($B=2$~T). The thick dashed red curve is the approximate formula $\left[ 1-\exp{\left(-\frac{\pi t^2 \tau_I}{\hbar \varepsilon_A}\right)}\right]\exp\left(-\frac{\pi \Delta E^2 \tau_I}{2\hbar \varepsilon_A}\right)$, obtained from the Landau-Zener probabilities, for $t=10\,\mu$eV and $B=0.5$~T. The thick solid blue curve is for $t=20\,\mu$eV and $B=1$~T. }
\label{fig:initialization}
\end{figure}

\subsubsection{Initialization}\label{sec:initializ}

We first discuss an initialization procedure into the $|-\rangle$ logical state by a detuning sweep starting from a large positive $\varepsilon_I$ (first pulse in Fig.~\ref{fig:pulses}). This method is usually more efficient than the initialization at $\varepsilon_A$ into $|+\rangle$ (which is the ground state), based on relaxation: due to the larger gap at positive detuning, the $|S(0,2)\rangle$ ground state can be prepared faster and with higher fidelity. An analogous procedure, with a detuning pulse from $\varepsilon_A$ to large positive values, allows one to read-out the $|\pm \rangle$ states via charge sensing.

Starting from the ground state at $\varepsilon_I$, it is straightforward to evaluate numerically the time-evolution of $|\psi(\tau)\rangle$. The probabilities of the logical states at time $\tau_I$ are given by:
\begin{equation}\label{Ppm}
P_\pm (\tau_I)=|\langle \pm |\psi(\tau_I)\rangle|^2.
\end{equation}
The fidelity $P_-(\tau_I)$ is plotted in Fig.~\ref{fig:initialization} as function of the initialization time $\tau_I$ (we have used a linear ramp in detuning as illustrated in Fig.~\ref{fig:pulses}).
Due to the anticrossing induced by tunneling around $\varepsilon=0$, a sufficiently long $\tau_I$ is necessary to guarantee adiabaticity in the two lowest energy branches. This time scale is given by $\tau_I \gtrsim  \hbar |\varepsilon_A|/t^2 $, as seen by the comparison to the Landau-Zener probability in Fig.~\ref{fig:initialization}. However, a decrease of fidelity is obtained at large $\tau_I$ which can be attributed to the presence of the anticrossing point at $\varepsilon_B$. In fact, the requirement of a fully diabatic evolution at the $\varepsilon_B$ anticrossing is violated at small sweeping rate (large $\tau_I$). To improve the maximum fidelity, it is necessary to have $t \gg \Delta E$, and a possible strategy shown in Fig.~\ref{fig:initialization} is to increase the external magnetic field $B$, since this leads to a suppression of $\Delta E$. 

If, on the other hand, we want to improve the initialization fidelity by retaining the same value of the $\Delta E $ (which determines the $\pi$-rotation time, as discussed in Sec.~\ref{spin_rotation_analytic}), an alternative strategy is to increase simultaneously $t$ and $B$, as exemplified in Fig.~\ref{fig:initialization}.  
It is seen that, by doubling both $B$ and $t$, $\Delta E$  and the long-$\tau_I$ decay of the two curves are left essentially unchanged. On the other hand,  the curve with larger $t$ shows a marked improvement at shorter $\tau_I$ and allows one to achieve a faster initialization with a higher maximum fidelity. 
Further improvement could be achieved by using an optimized pulse shape instead of a simple linear ramp. For example,  if the two anticrossing regions $|\varepsilon| \lesssim t$ and  $|\varepsilon-\varepsilon_B| \lesssim \Delta E$ are well separated, a pulse with a different rate $d\varepsilon/d\tau$ in each region\cite{Ribeiro2013a} could be helpful to improve the fidelity.

\begin{figure}
\includegraphics[width=0.41\textwidth]{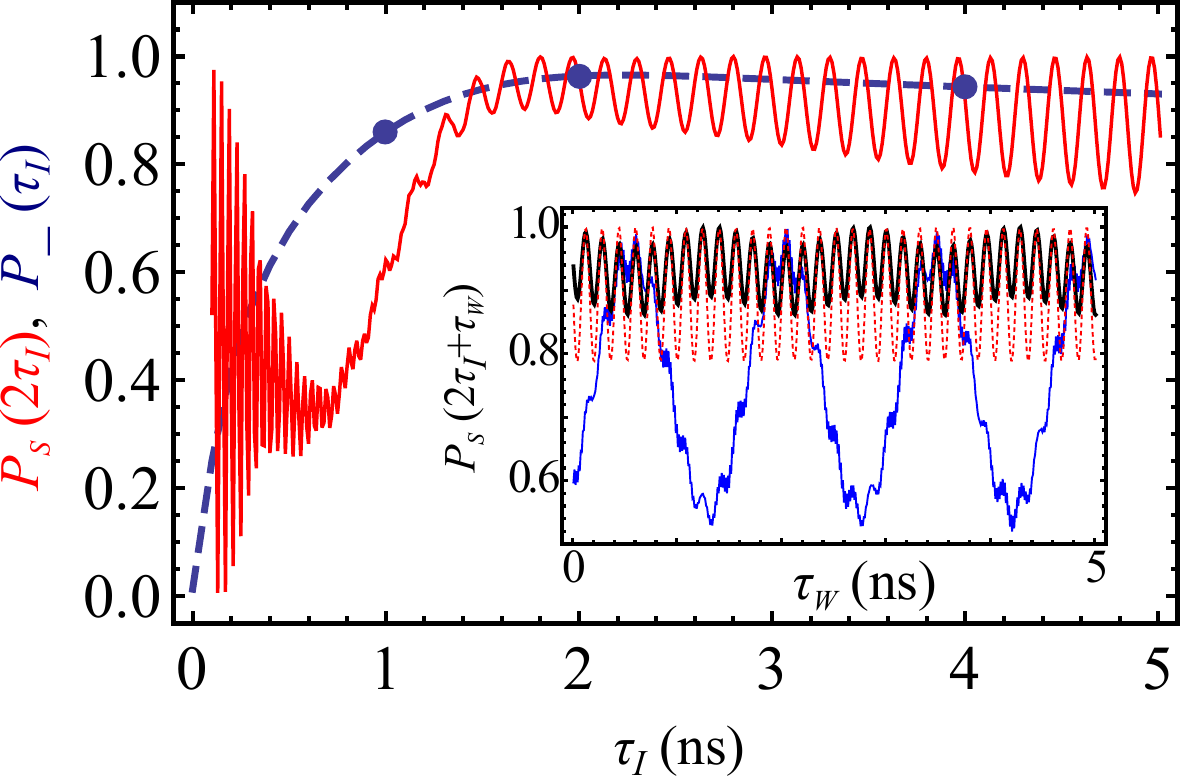}
\caption{Main panel: The solid red line is the return probability $P_S(2\tau_I)$ to the $\varepsilon(\tau=0)=200~\mu$eV ground state after two linear detuning ramps of duration $\tau_I$: first to $\varepsilon(\tau_I)=-200~\mu$eV and than back to $\varepsilon(2\tau_I)=200~\mu$eV.  We used $t=10\,\mu$eV, and the geometry of Fig.~\ref{fig:micromagnet}(b) with $\varphi=0$, $B=1$ T. The growing oscillations at large $\tau_I$ reflect the larger amplitude of the $|+\rangle$ state at the initialization time $\tau_I$.  The dashed blue line shows $P_-(\tau_I)$, reproduced from Fig.~\ref{fig:initialization}. The inset shows the return probability by introducing a waiting time $\tau_W$ between the two linear ramps. The thin blue, thick black, and dotted red curves correspond respectively to $\tau_I=1,2,4$~ns (see dots in the main panel). A smaller oscillation amplitude is obtained for $\tau_I=2$ ns.}
\label{initialization2}
\end{figure}

Finally, it is worth noting that, as far as the unitary evolution is concerned, the fact that $P_-(\tau_I)<1$ does not pose a significant problem if initialization in the eigenstates of $\tilde{S}_{2,z}$ is not required. Rather, a longer $\tau_I$ allows to achieve a higher initialization fidelity into the logical subspace but the initialization axis (and readout axis as well, by considering the inverse detuning pulse) will be tilted with respect to the logical $\hat z$. The tilt angle can be made larger with a longer $\tau_I$, before dephasing mechanisms become relevant. Initialization in a superposition of eigenstates can be addressed experimentally with a pulse which returns to large positive detuning, to readout the $S(0,2)$ probability through charge sensing.\cite{Petta2010,Ribeiro2013b,Wu2014} The resulting quantum oscillation are illustrated in Fig.~\ref{initialization2}.

\subsubsection{Single-spin rotations}

As $z$-rotations can be easily implemented through the Zeeman splitting (see the discussion at the beginning of Sec.~\ref{sec_analysis}), we focus here exclusively on the spin manipulation realized using the anticrossing at $\varepsilon_B$. The detuning pulse, with total gate time $\tau_G$, is illustrated in Fig.~\ref{fig:pulses} and we show numerical results with the system initialized in the $|-\rangle$ state. This is sufficient to illustrate the gate performance if the total fidelity $F(\tau_G)=P_+(\tau_G)+P_-(\tau_G)$ is close to 1, thus a nearly unitary operation is realized within the logical subspace [$P_\pm(\tau_G)$ are defined as in Eq.~(\ref{Ppm})].  

\begin{figure}
\includegraphics[width=0.48\textwidth]{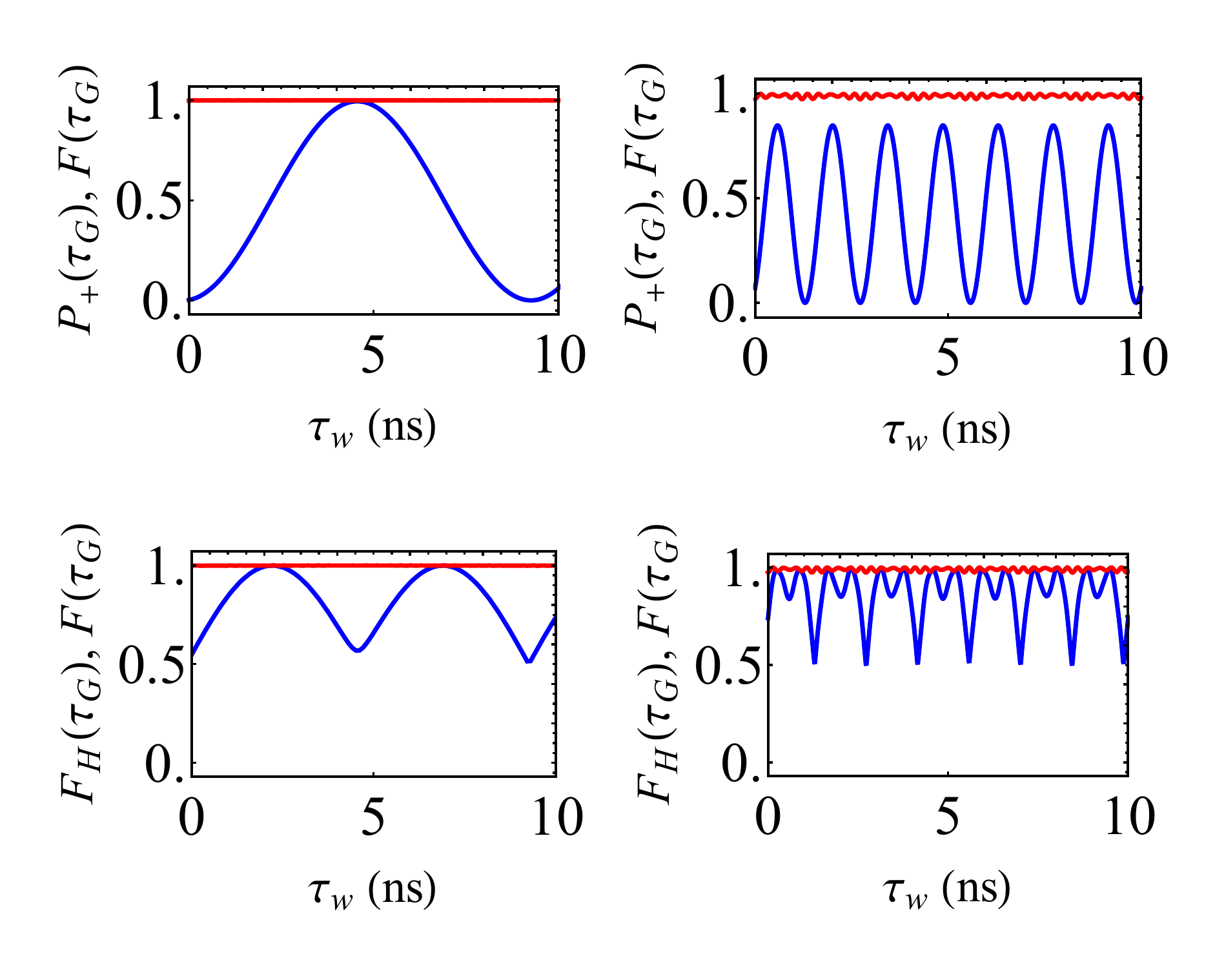}
\caption{
Upper panels: the blue lower curves show $P_+(\tau_G)$ as function of $\tau_W$ (i.e., the waiting time at $\varepsilon_B $ such that $\tau_G=\tau_W+2\tau_R$, see Fig.~\ref{fig:pulses}). The left panel is for the setup in Fig.~\ref{fig:micromagnet}(a) with $\varphi=5\pi/8$ and the right panel is for Fig.~\ref{fig:micromagnet}(b) with $\varphi=0$. Lower panels: the blue lower curves show $F_H(\tau_G)$ as defined in Eq.~(\ref{FH}) for the same parameters as the upper panels (as above, left and right panels are for the setups of Fig.~\ref{fig:micromagnet}(a) and (b), respectively). In all panels the upper red curves are $F(\tau_G)$, $t=25~\mu$eV, $B=0.5$ T, $\tau_R=0.2$ ns, and the initial state is $|-\rangle$ at $\varepsilon_A = -200 \, \mu$eV.
}
\label{fig:XHgates}
\end{figure}

\begin{figure}
\includegraphics[width=0.4\textwidth]{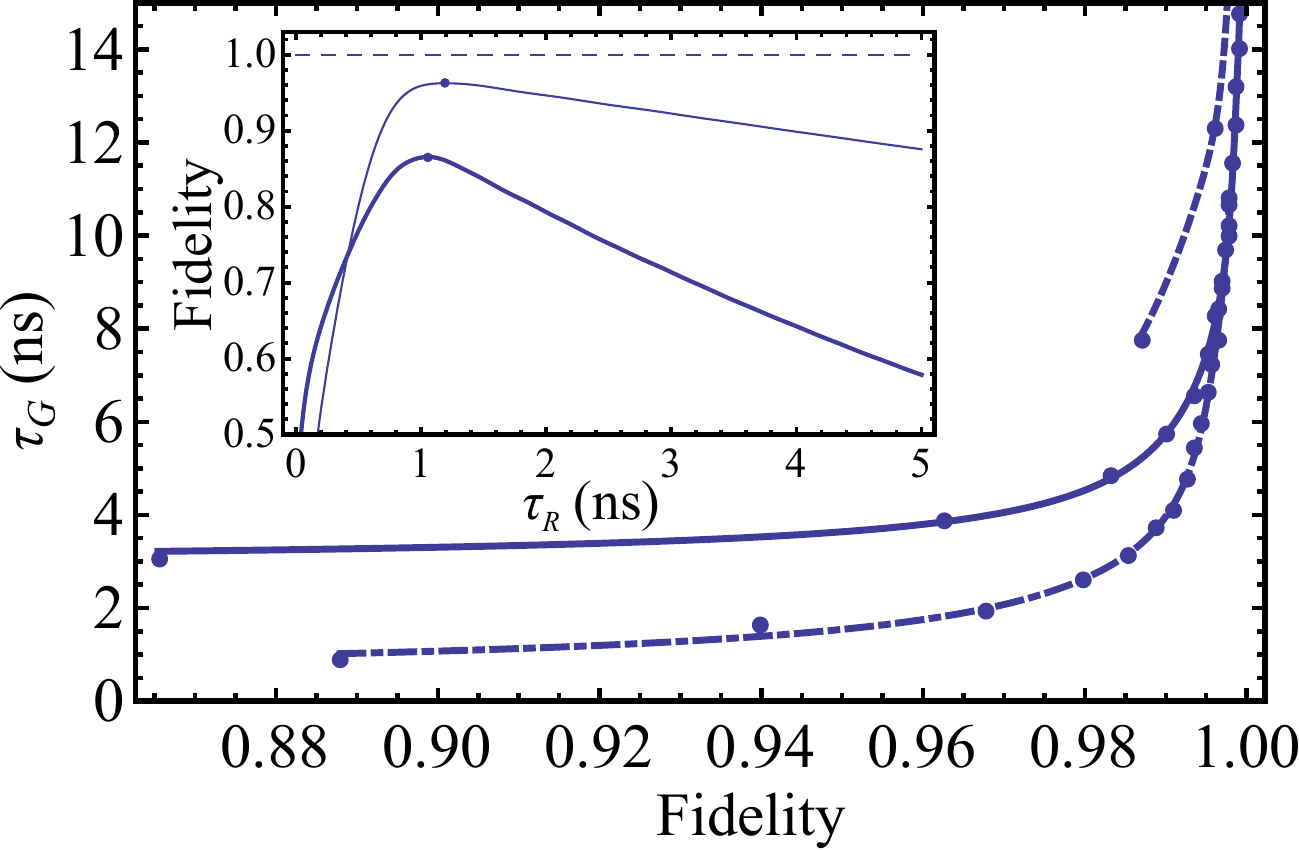}
\caption{Main panel: Maximum value of $P_+(\tau_G)$ obtained with an optimum value $\tau_R^*$, the initial state $|-\rangle$ ($\varepsilon_A = -200 \, \mu$eV), and a waiting time $\tau_W=\tau_\pi$ at $\varepsilon_B$, see Eq.~(\ref{tauPi}). The plot illustrates the relation between the $\pi$-rotation fidelity $P_+(\tau_G)$ and the corresponding total gate time $\tau_G=\tau_\pi+2\tau_R^*$. The three curves are a guide for the eye through the numerical data (dots), computed for $B=0.5,1,1.5,\ldots$ T from left to right.  The dashed curve refers to the geometry of Fig.~\ref{fig:micromagnet}(a) with $\varphi=5\pi/8$ and $t=10\,\mu$eV. The other two curves are for Fig.~\ref{fig:micromagnet}(b) with $\varphi=0$ and $t=10\,\mu$eV (solid) or $t=20\,\mu$eV (dot-dashed). The inset shows two examples on how to determine the optimum value $\tau_R^*$ from the maximum in $P_+(\tau_G)$ as function of $\tau_R$. The slanting field is the same as in Fig.~\ref{fig:micromagnet}(b), $t=10\,\mu$eV, and the thick (thin) curve is for $B=0.5\,{\rm T}$ ($1\,{\rm T}$).}
\label{fig:xgate}
\end{figure}

As expected from the discussion in Sec.~\ref{spin_rotation_analytic}, the detuning pulse should realize a rotation about an axis perpendicular to $\hat z$, thus allowing one to implement a $\pi$-rotation equivalent to a NOT-gate. The two upper plots in Fig.~\ref{fig:XHgates} show the behavior of $P_+(\tau_G)$ for two representative cases. As expected, $P_+(\tau_G)$ displays oscillations in $\tau_G$ with period $2\tau_\pi$, which are significantly faster for the setup (b) of Fig.~\ref{fig:micromagnet} than for setup (a). We obtain $P_+(\tau_G)\simeq 1$ for suitable parameters. Trying to optimize the fidelity of the $\pi$-rotation, we find a non-monotonic dependence on $\tau_R$ similar to the optimization of the initialization fidelity with respect to $\tau_I$. An example is shown in the inset of Fig.~\ref{fig:xgate}: the maximum in fidelity occurs at $\tau_R^*$, which is determined by the competition of the two relevant anticrossings (at $\varepsilon=0$ and $\varepsilon_B$) in requiring an adiabatic/diabatic evolution within the lower two energy branches. Similarly as before, an increase in the external field $B$ leads to higher values of the fidelity due to the suppression of $\Delta E$, thus to a better energy scale separation $t \gg \Delta E$. However, a larger $B$ also degrades the gate time $\tau_G=\tau_\pi+2\tau_R^*$ due to the longer $\tau_\pi$. 

To clarify the typical interplay between relevant parameters, we show in Fig.~\ref{fig:xgate} the relation between the optimum fidelity of a $\pi$-rotation and the corresponding gate time $\tau_G$. As the external field $B$ is increased, a better fidelity approaching 1 is obtained at the expense of a longer $\tau_G$. In the geometry of Fig.~\ref{fig:micromagnet}(b) the same fidelity of setup (a) can be achieved with a shorter gate time, as seen by a comparison between the solid and dashed curves of Fig.~\ref{fig:xgate}. The gate time can be further improved if the tunneling energy is made larger, as seen by a comparison of the solid and dot-dashed curves of Fig.~\ref{fig:xgate}. 

The reduced fidelity of the $\pi$-rotation in the favorable regime of larger values of $\Delta E$ (and shorter gate times) does not prevent in general to achieve effective spin-manipulation since we obtain $F(\tau_G)\simeq 1$ in Fig.~\ref{fig:XHgates}. Thus, the smaller maximum value of $P_+(\tau_G)$ (see the top right panel of Fig.~\ref{fig:XHgates}) can be simply attributed to a rotation axis which is not perpendicular to $\hat{z}$, due to the imperfect realization of the diabatic evolution Eq.~(\ref{ideal_sweep}). In particular, when $P_+(\tau)=1/2$ a $\pi/2$ rotation from $\hat z$ to the $xy$-plane (equivalent to an Hadamard gate) is realized with high accuracy. We can characterize the fidelity of this $\pi/2$-rotation through the in-plane spinors $|\phi\rangle=\left(|+\rangle+e^{-i\phi}|-\rangle\right)/\sqrt{2}$ (with $\phi$ an arbitrary phase). Choosing $\phi$ to maximize the overlap with $|\psi(\tau)\rangle$, we obtain for the Hadamard gate:
\begin{equation}\label{FH}
F_H(\tau_G)\equiv {\rm max}_\phi|\langle \phi |\psi(\tau_G)\rangle|^2=\frac12\left(\sum_\pm\sqrt{P_\pm(\tau_G)}\right)^2.
\end{equation}
This quantity is plotted in the two lower panels of Fig.~\ref{fig:XHgates} and is simply related to $P_+$ of the upper panels by the approximate relation $F_H\simeq 1/2+\sqrt{P_+(1-P_+)}$ (using $F\simeq 1$). Thus,
 $F_H(\tau_G)$ approaches one when $P_+(\tau_G)=1/2$ and is bounded by the total fidelity $F(\tau_G)$. The NOT-gate can be alternatively realized by making use of a composition of $z$-rotations and two of these $\pi/2$-rotations. As the unitary dynamics allows for high-fidelity universal control, it becomes important to consider limitations introduced by relevant dephasing mechanisms, which are discussed in the next section.

\subsection{Decoherence mechanisms}\label{sec:noise}

We estimate here the decoherence timescales and the expected analytic form of decay induced by the hyperfine interaction and charge noise. From the resulting parameter dependence, we suggest under what conditions these decoherence effects can be made small. 

\subsubsection{Hyperfine interaction}\label{sec:hyperfine}

\begin{figure}
\includegraphics[width=0.4\textwidth]{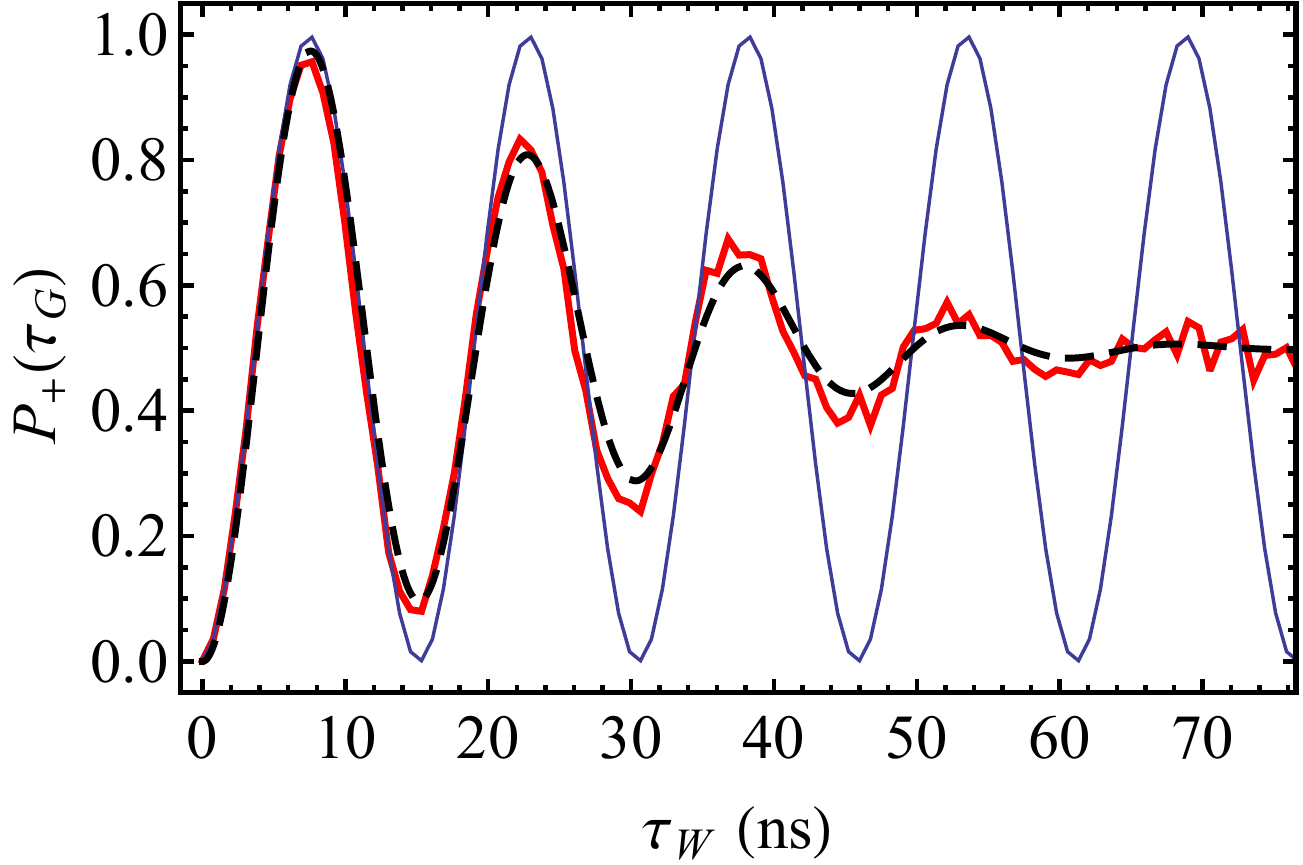}
\caption{Plot of the fidelity $P_+(\tau_G)=|\langle + |\psi(\tau_G)\rangle|^2$ for a rotation starting from $|\psi(0)\rangle=|-\rangle$ ($\varepsilon_A = -200
\, \mu$eV) as a function of the waiting time $\tau_W$ at $\varepsilon_B$. We used $\tau_R=2\,{\rm ns}$, $t=10\,\mu$eV, and the geometry of Fig.~\ref{fig:micromagnet}(a) with $B=1\,{\rm T}$ and $\varphi=5\pi/8$. The thin solid line is obtained without nuclear dephasing, while the thick red curve includes nuclear fluctuations with $\sigma=2\,{\rm mT}$ (average over 400 runs). The thick black dashed line  is Eq.~(\ref{gaussian_decay}) with $\tau_N$ given by Eq.~(\ref{tauN}).}
\label{nuclei_decay}
\end{figure}

As the $\pi$-rotations can be realized on a rather short time scale $\ll 10\,{\rm ns}$, see Figs.~\ref{fig:XHgates} and \ref{fig:xgate}, it becomes justified to approximate the nuclear environment with static random fields, which modify the values of ${\bf b}_{1,2}$. Also a recently discussed nuclear dephasing mechanism, induced by the inhomogeneous magnetic field,\cite{Beaudoin2013} becomes only relevant at much longer times. We consider nuclear fields which are uncorrelated between the two dots and have a Gaussian probability distribution with zero mean and variance $\sigma^2$ for each component, as discussed in previous works.\cite{Coish2004,Coish2007} We have computed the result in Fig.~\ref{nuclei_decay} for a particular set of parameters, where it can be seen that the effect on the fidelity at the first maximum is small. In fact, fluctuations induced by the nuclei are of order $\sigma\sim 1\,{\rm mT}$. Since it is possible to realize $\Delta b \simeq 100\,{\rm mT}$ through the slanting field of the micromagnet, the effect of the nuclei on the anticrossing can be very small. 

To obtain a quantitative expression, we assume that the evolution between $\varepsilon_{A,B}$ is realized as in Eq.~(\ref{ideal_sweep}). Since the random change in $\Delta E$ is small, it is justified to consider only the linear correction from the nuclear field. In this approximation $\Delta E$ has a Gaussian distribution, which yields the following expression for $P_+(\tau_G)$:
\begin{equation}
P_+(2\tau_R+\tau)\simeq \frac{1}{2}\left[1-e^{-(\tau/\tau_N)^2}\cos{(\overline{\Delta E} \tau/\hbar)}\right].
\label{gaussian_decay}
\end{equation}
The overline indicates the average over nuclear configurations and $\tau_N^2=2\hbar^2/(\overline{\Delta E^2}-\overline{\Delta E}^2)=2\hbar^2/\sigma_{\Delta E}^2$. To estimate $\overline{\Delta E}$, we can simply use the unperturbed values of $\theta,b$ in Eq.~(\ref{DeltaE}). We also obtained to lowest order in $\sigma^2$:
\begin{equation}
\sigma_\theta^2 =\left( \frac{1}{b_1^2}+\frac{1}{b_2^2}\right) \sigma^2 ,\qquad  \sigma_b^2 =\frac{\sigma^2}{2},
\end{equation}
while the covariance is zero to the same order of approximation. From these results, $\sigma_{\Delta E}^2$ can be obtained from Eq.~(\ref{DeltaE}) as usual:
\begin{equation}
\sigma_{\Delta E}^2 = \left(\frac{\partial \Delta E}{\partial \theta}\right)^2\sigma_\theta^2+\left(\frac{\partial \Delta E}{\partial b}\right)^2\sigma_b^2,
\end{equation}
which is easily evaluated and yields results in good agreement with the numerical evaluation. 

A relevant regime which is more transparent to discuss is when $\Delta E \simeq t \theta$, see Eq.~(\ref{DeltaE_approx}). In this case the fluctuations in $\Delta E$ are directly related to the fluctuations in the angle $\theta$. By using $\sigma_\theta^2\simeq 2\sigma^2/b^2$, we obtain the following decay time scale:
\begin{equation}
\tau_N \simeq \frac{\hbar b}{\sigma t},
\label{tauN}
\end{equation}
which, using $\theta\simeq \Delta b_\perp/b$, can be compared to the characteristic gate time from Eq.~(\ref{tauPi}):
\begin{equation}
\tau_\pi \simeq \frac{\pi \hbar b}{\Delta b_\perp t}.
\label{taupi_approx}
\end{equation}
In Fig.~\ref{nuclei_decay} we have used Eq.~(\ref{tauN}), together with Eq.~(\ref{gaussian_decay}), and obtained a satisfactory description of the decaying oscillations. An interesting feature of Eq.~(\ref{tauN}) is that for this problem the relevant nuclear dephasing timescale $\tau_N$ is proportional to $b\simeq B$. This is easily understood since $\Delta E \simeq t \theta$ and the typical change in the angle $\theta$ due to the nuclear field is $\delta \theta_N \sim\sigma/b$, thus the fluctuations in $\theta$ (and in $\Delta E$) are suppressed by a larger value of $b$. Similarly, a larger magnetic field will increase the oscillation period, which is also proportional to $b$. Therefore, the `quality factor' remains constant since the ratio $\tau_N /\tau_\pi  =\Delta b_\perp/(\pi\sigma)$ is independent of $b$. In particular, the ratio of timescales is governed by $\Delta b_\perp/\sigma$, where $\Delta b_\perp$ is the difference in local fields transverse to ${\bf B}$, see Eq.~(\ref{b_perp}). As discussed, this ratio can be made large since $\Delta b_\perp$ can be of order $\sim 100\,{\rm mT} \gg \sigma \sim 1\,{\rm mT}$. In conclusion, these arguments indicate that the spin manipulation scheme can be rather insensitive to the influence of the nuclear bath.

\subsubsection{Charge noise}

\begin{figure}
\includegraphics[width=0.45\textwidth]{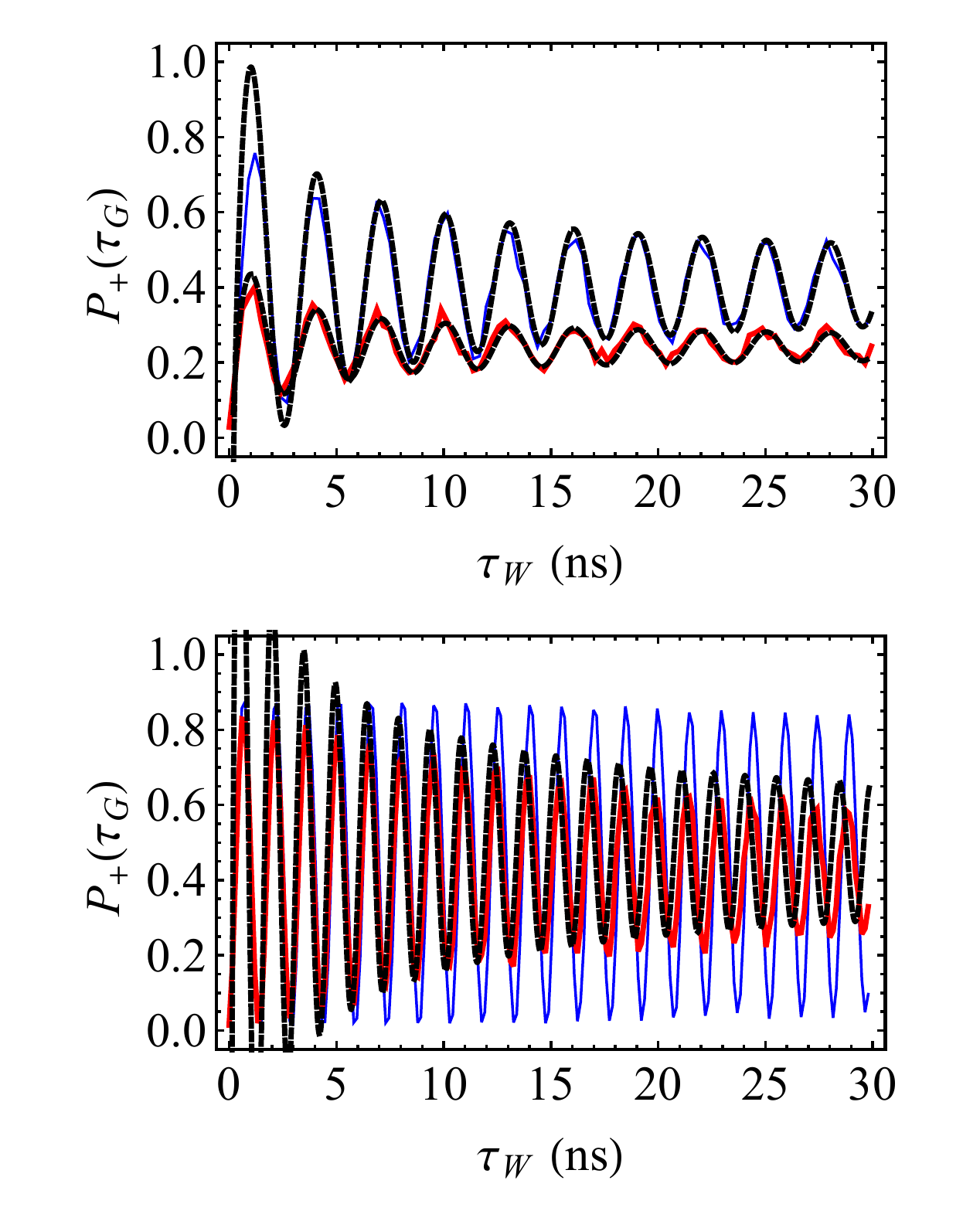}
\caption{
Plot of the fidelity $P_+(\tau_G)=|\langle + |\psi(\tau_G)\rangle|^2$ for a rotation of $|-\rangle$ ($\varepsilon_A = -200 \, \mu$eV), as function of the waiting time $\tau_W$ at $\varepsilon_B$. For both panels we assumed the slanting field of Fig.~\ref{fig:micromagnet}(b) with $\varphi=0$. The solid curves are averages of 400 runs using $\sigma_\varepsilon=1~\mu$eV for the thin blue curves and $\sigma_\varepsilon=3~\mu$eV for the thick red curves. In the upper panel $\tau_R=2$ ns, $B=1$ T, and $t=10~\mu$eV, which give $\Delta E=1.4~\mu{\rm eV}$. In the lower panel $\tau_R=0.1$ ns, $B=0.5$ T, and $t=20~\mu$eV, which give $\Delta E=2.8~\mu{\rm eV}$. The dashed curves refer to the asymptotic expression Eq.~(\ref{power_law_decay}). The $\sigma_\varepsilon=1~\mu$eV curve of the second panel has small decay and Eq.~(\ref{power_law_decay}) is not plotted, as it is applicable only for $\tau_W \gg \hbar\Delta E/ {\sigma'_{\varepsilon}}^2 \simeq 35$ ns.\cite{Koppens2007} }
\label{fig:chargenoise}
\end{figure}

As recent experiments have demonstrated the important role played by low-frequency charge fluctuations,\cite{Dial2013,Kornich2014} we consider the effect of charge noise on the single-spin rotations. We introduce a random shift $\delta\varepsilon$ in detuning, assuming a Gaussian distribution with variance $\sigma_{\varepsilon}$. This type of noise displaces the operating points from the desired values. Especially, the $\pi$-rotations are now realized at a detuning  $\varepsilon_{B}+\delta \varepsilon$  which does not coincide with the anticrossing point. A certain degree of protection against dephasing arises from the fact that $\varepsilon_B$ is a stationary point for the energy gap and, to lowest order, the change in the gap energy $\Delta E$ is quadratic in $\delta \varepsilon$. The relevant scale for $\sigma_\varepsilon$ is set by $\Delta E$. Figure~\ref{fig:chargenoise} shows a strong suppression of the visibility when $\sigma_\varepsilon \gtrsim \Delta E$, while the coherent oscillations are significantly more robust when $\sigma_\varepsilon \lesssim \Delta E$. 

A more precise description of this effect can be obtained by noticing, from the effective model in Eq.~(\ref{H_eff}), that the noise in $\varepsilon$ induces a perturbation along the effective $z$-direction. The variance of this perturbation is obtained from:
\begin{equation}
{\sigma'_{\varepsilon}}=\frac12 \left(1+\left.\frac{\partial \Delta}{\partial \varepsilon}\right|_{\varepsilon_B}\right)\sigma_\varepsilon =\frac{\sigma_\varepsilon}{1+2\left(\frac{t\cos (\theta /2)}{|g|\mu
_{B}b}\right)^2}. \label{sigma_eff}
\end{equation}
On the other hand, fluctuations in the off-diagonal element of Eq.~(\ref{H_eff}) can be readily calculated to linear order in $\delta\varepsilon$, which gives:
\begin{eqnarray}
{\sigma''_{\varepsilon}}&& =\left( \left .\frac{\partial }{\partial \varepsilon}\sqrt{\frac{\Delta +\varepsilon }{2\Delta }} \right|_{\varepsilon_B}t\sin (\theta /2)\right)\sigma_\varepsilon \nonumber \\
&& =\frac{2 \tan(\theta/2)}{\left[2+\left(\frac{|g|\mu
_{B}b}{t\cos ( \theta /2)}\right)^2 \right]^{3/2}}{\sigma'_{\varepsilon}} < \frac{\tan(\theta/2)}{\sqrt{2}}{\sigma'_{\varepsilon}}. 
\label{sigma_eff2}
\end{eqnarray}
We see that typically ${\sigma''_{\varepsilon}}/{\sigma'_{\varepsilon}}\ll 1$, due to the small angle $\theta$ (an additional small factor appears for $t \ll |g| \mu_B b$). If we neglect the effect of gate noise in the off-diagonal element, the problem is formally equivalent to the theory of Ref.~\onlinecite{Koppens2007} describing the decay of Rabi oscillations due to the transverse fluctuations of the Overhauser field. This correspondence yields the following asymptotic result:
\begin{eqnarray}
P_+(\tau)\simeq && \frac{\sqrt{2\pi}\Delta E}{4{\sigma'_{\varepsilon}}}\exp{\left(\frac{\Delta E^2}{2{\sigma'_{\varepsilon}}^2}\right)} {\rm erfc}\left(\frac{\Delta E}{\sqrt{2}{\sigma'_{\varepsilon}}}\right) \nonumber \\
&& -\frac{1}{2}\sqrt{\frac{\Delta E \hbar}{{\sigma'_{\varepsilon}}^2 \tau}}\cos{\left(\Delta E \tau/\hbar+\frac{\pi}{4}\right)},
\label{power_law_decay}
\end{eqnarray}
with ${\rm erfc}(x)$ the complementary error function. This expression is characterized by a power-law decay and a universal $\pi/4$ phase shift. As seen in the first panel of Fig.~\ref{fig:chargenoise}, Eq.~(\ref{power_law_decay}) is able to reproduce accurately the asymptotic form of the coherent oscillations. For larger values of $t$ and smaller $B$, as in the second panel of Fig.~\ref{fig:chargenoise},  several assumptions in deriving the simple form of Eq.~(\ref{power_law_decay}) become less accurate: $\theta$ is larger (leading to an increase of  ${\sigma''_{\varepsilon}}/{\sigma'_{\varepsilon}}$), higher-order corrections to the $H_{\rm eff}$ of Eq.~(\ref{H_eff}) become more relevant, and even without charge noise the amplitude of the $P_+(\tau_G)$ oscillations is significantly smaller than one. Nevertheless, the power-law decay is still in qualitative agreement with the numerical results. The asymptotic formula Eq.~(\ref{power_law_decay}) is valid for:\cite{Koppens2007}
\begin{equation}\label{tau_polynomial}
\tau \gg  \frac{\Delta E \hbar}{{\sigma'_{\varepsilon}}^2},
\end{equation}
when $\Delta E /{\sigma'_{\varepsilon}}>1$. Equation (\ref{tau_polynomial}) also provides the relevant time scale for a significant reduction in visibility due to the $\tau^{-1/2}$ prefactor, see the second line of Eq.~(\ref{power_law_decay}), thus gives a quality factor $\tau/\tau_\pi \sim (\Delta E/{\sigma'_{\varepsilon}})^2$. The value of ${\sigma'_{\varepsilon}}$ could be effectively reduced if $t^2/|g\mu_B B|^2\gg 1$ in the denominator of Eq.~(\ref{sigma_eff}). In fact, there is no fundamental limitation in our simple model to reduce charge noise by increasing $t$. In other words, our scheme for rotations at $\varepsilon_B$ is based on the $|\tilde{S}_- \rangle $ state of Eq.~(\ref{Stilde_pm}). While $|\tilde{S}_- \rangle $ necessarily introduces a superposition of different spin states, it is still possible to suppress in Eq.~(\ref{Stilde_pm}) the amplitude of $|\tilde{S}(0,2)\rangle$ through a large value of $t$, which makes this state closer to a pure (1,1) charging configuration, thus less sensitive to fluctuations in $\varepsilon$. 

\begin{figure}
\includegraphics[width=0.4\textwidth]{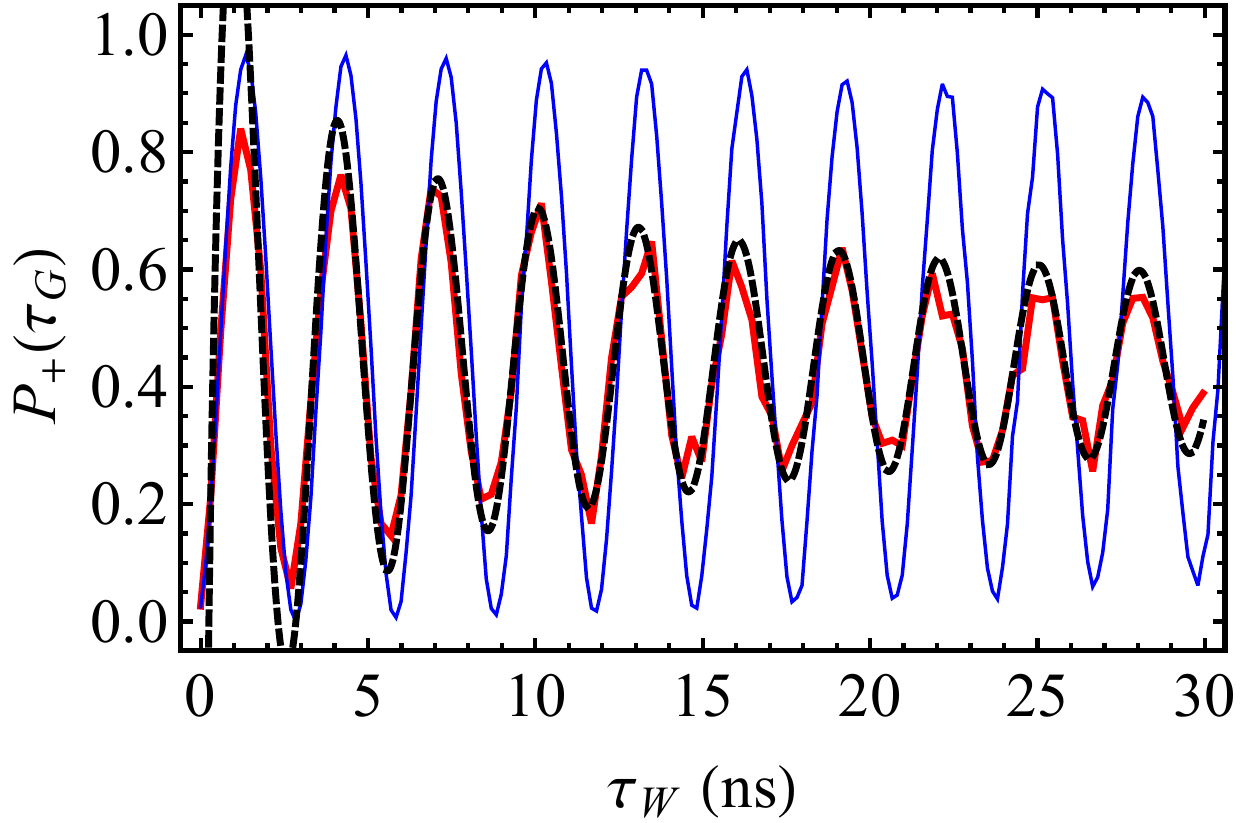}
\caption{
Plot of the fidelity $P_+(\tau_G)=|\langle + |\psi(\tau_G)\rangle|^2$ with the same parameters of the upper panel of Fig.~\ref{fig:chargenoise}, but including fluctuations in the tunnel amplitude $t$ instead of $\varepsilon$. The thick red (thin blue) solid curve is for $\sigma_t/t =5\%$ ($\sigma_t/t =1\%$). The dashed curve is the asymptotic formula for the $\sigma_t/t =5\%$ curve, calculated using Eq.~(\ref{sigma_eff_t}). 
}
\label{fig:tunnelnoise}
\end{figure}

Besides random shifts of $\varepsilon$, static fluctuations on the tunneling barrier can also be simply treated by introducing a Gaussian variation of $t$, with variance $\sigma_t^2$. Similar as detuning noise, we obtain that the noise fluctuations in the effective Hamiltonian Eq.~(\ref{H_eff}) are prevalently along the effective $z$-direction. This leads to the same expression Eq.~(\ref{power_law_decay}) with ${\sigma'_{\varepsilon}}$ replaced by the corresponding ${\sigma'_{t}}$, obtained as follows:
\begin{equation}
{\sigma'_{t}}=\frac12 \left.\frac{\partial \Delta}{\partial t}\right|_{\varepsilon_B} \sigma_t =
\frac{2 \sigma_t \cos (\theta /2)}{\frac{|g|\mu_{B}b}{2t\cos (\theta /2)}+\frac{t\cos (\theta /2)}{|g|\mu_{B}b}}. \label{sigma_eff_t}
\end{equation}
As seen in Fig.~\ref{fig:tunnelnoise}, the agreement with the numerics is good. As a result, although the parameter dependence is different, it might be difficult to distinguish the two possible effects of charge noise (tunnel and detuning fluctuations) from the asymptotic form of the coherent oscillations. For example, the decoherence of Fig.~\ref{fig:chargenoise} could also be interpreted as due to noise in $t$ such that $\sigma_{t}'=\sigma'_\varepsilon$. By combining Eqs.~(\ref{sigma_eff}) and (\ref{sigma_eff_t}) we get that the $\sigma_\varepsilon =1,3~\mu$eV curves of the upper panel of Fig.~(\ref{fig:chargenoise}) would correspond to $\sigma_t/t \simeq 7\%$, $21\%$ while for the lower panel $\sigma_t/t \simeq 1\%$, $3\%$. On the other hand, as discussed in Sec.~\ref{sec:hyperfine}, static nuclear fluctuations give rise to a distinct form of gaussian decay.

\section{Discussion and Conclusions}\label{sec:conclusion}

We have characterized a spin manipulation scheme involving the two lowest energy states of a double quantum dot in the slanting field of a micromagnet. Working at sufficiently large negative detuning, this scheme effectively realizes single-spin rotations in one of the two quantum dots, with the other dot serving as an auxiliary spin. The general principle of operation is similar to Ref.~\onlinecite{Coish2007} but the physical picture is different: the auxiliary spin of Ref.~\onlinecite{Coish2007} is ``pinned'' by a large local field $b_{1}\gg b_{2}$ and its role is to induce through the exchange interaction an effective local field (parallel to ${\bf b}_1$) on the second spin; instead, in our case we have $b_1\simeq b_2 \simeq B$ and at the $\varepsilon_B$ anticrossing the two spins become strongly entangled. 

In our parameter regime, fast spin manipulation ($\sim 1 $ ns) can be achieved without requiring:\cite{Coish2007}
\begin{equation}
b_{2,x} \ll  b_{2,z} \ll b_{1,z}, \label{b_condition_bill}
\end{equation}
a condition which in practice can turn to be too restrictive. In fact, the timescale of $x$-rotations given in Ref.~\onlinecite{Coish2007} is $\hbar/|g\,\mu_B b_{2,x}|$ (as in Sec.~\ref{local_spins} we choose ${\bf b}_1$ along $z$ and $b_{2,y}=0$). Since Eq.~(\ref{b_condition_bill}) implies $ b_{1,z}\simeq  b_{1,z}-b_{2,z}$, it is difficult to realize $b_{1,z}$ much larger than $ \simeq 100$ mT. Therefore, if Eq.~(\ref{b_condition_bill}) is strictly enforced in GaAs lateral quantum dots, $b_{2,x} $ could become comparable to the $\sigma \sim 1$ mT nuclear field fluctuations  

While several strategies were proposed to realize Eq.~(\ref{b_condition_bill}),\cite{Coish2007} we have found that an alternative method is to simply maximize $b_{2,x}$ (say, $b_{2,x}\sim 100$ mT) while satisfying:
\begin{equation}
b_{2,x} \ll  b_{2,z} \simeq b_{1,z}, \label{b_condition_us}
\end{equation}
which can be always realized with a sufficiently strong external field ($b_{i,z}\simeq B$). In this case, the limiting operation time $\sim \hbar/|g\,\mu_B b_{2,x}|$ is approached when $t \simeq |g|\mu_B B$, see Eq.~(\ref{tauPi_approx}). Thus in our case a relatively large tunneling amplitude is favorable. A large tunneling amplitude is also useful to suppress the effect of fluctuations in detuning, when $t^2/|g\mu_B B|^2 \gg 1$. Effectively, this scheme takes advantage of the large energy scales set by $t$ and $|g|\mu_B B$, to achieve an improved fidelity and operation time. An obvious limitation where this strategy breaks down is given by the orbital energy scale of the quantum dots, but this is $\sim 1$ meV for GaAs lateral dots (see, e.g., Refs.~\onlinecite{Nowack2007,Pioro2008}). The effect of nuclear spins depends on the $\sigma/b_{2,x}$ ratio, which is typically small in the optimal regime. Thus, this approach could realize high fidelity single-spin gates on a timescale of $\sim 1$ ns.

A related method for spin manipulation which was experimentally demonstrated makes use of Landau-Zener-St\"uckelberg interferometry through the $\varepsilon_B$ anticrossing, induced in that case by the nuclear fields.\cite{Petta2010,Ribeiro2013a,Ribeiro2013b} Landau-Zener interferometry yields an alternative approach for the manipulation of $|\pm \rangle$. The gate time would be determined by the same timescale $\hbar/\Delta E$ discussed so far, and the use of a micromagnet slanting field should allow for significant improvements over nuclear fields. A relevant process discussed in those works is the phonon-mediated relaxation at $\varepsilon_B$, which yields a slow timescale $\Gamma^{-1} \sim 10~\mu$s when $\Delta E \sim k_B T$, by fitting a phenomenological model.\cite{Ribeiro2013a} However, spin relaxation processes mediated by phonons can have a strong dependence on the gap $\Delta E$ and this estimate might not be appropriate in our case. A more detailed microscopic theory\cite{Khaetskii2001,Golovach2004, Stano2006,Amasha2008,Kornich2014,Scarlino2014} would be necessary to assess this effect. We have also neglected spin-orbit coupling terms, which have an effect on the anticrossing with a complicated dependence on the double dot parameters.\cite{Stepanenko2012} As their energy scale is comparable to the nuclear field fluctuations, we expect a small influence on our discussion of $\Delta E$ and spin manipulation.

In concluding, we stress again that the qubit is encoded here into the single-spin states of one of the dots, even if a double dot is used for spin manipulation. Thus, two-qubit gates could be implemented by simply controlling the exchange interaction between two target spins,\cite{Loss1998,Petta2005,Coish2007} which is a potential advantage with respect to other types of encoding using multiple quantum dots.

\acknowledgments
We thank W. A. Coish, S. N. Coppersmith, M. Delbecq, and P. Stano for helpful discussions. 
We acknowledge support from the IARPA project 
“Multi-Qubit Coherent Operations”
through Copenhagen University. S.C. and Y.D.W. acknowledge
support from the 1000 Youth Fellowship Program
of China. J.Y., T.O., and S.T. acknowledge support from
ImPACT Program of Council for Science Technology and
Innovation (Cabinet Office, Government of Japan), Grants-in-Aid 
for Scientic Research S (No. 26220710), and FIRST.
T.O. acknowledges support from the Japan Prize Foundation,
JSPS, RIKEN incentive project, and the Yazaki Memorial
Foundation. D.L. acknowledges support from the Swiss NSF
and NCCR QSIT.

\appendix

\section{Corrections to the factorized form of the logical states}\label{appendix1}

In this Appendix, we discuss the leading corrections to Eq.~(\ref{factorized}). For $| -
\rangle$ the probability of admixture with $|\tilde S(0,2)\rangle$,
introducing undesired entanglement between the two spins, is of order $%
t^2/\varepsilon_A^2 \sim 0.06 \%$, using realistic values of $t = 5\,\mu$eV
and $\varepsilon_A=-200\,\mu$eV, and can be systematically reduced by
increasing $|\varepsilon_A|$ (while to keep a relatively large value of $t$
is desirable). $|+\rangle \simeq |\psi_{+,+}\rangle$ has a much
larger purity since typically $\sin (\theta/2) \ll 1$. For the $|-\rangle$
logical state, we estimate that the probability of mixing with $|
\psi_{-,+}\rangle$ is of order $t^4/(\varepsilon_Ag\mu_B\Delta b)^2$. Since
a realistic magnetic field gradient gives $|g\mu_B\Delta b |\sim 1 \,\mu$eV,
the factor $t^2/(g\mu_B\Delta b)^2$ contributes to enhance the admixture
fraction with respect to $t^2/\varepsilon_A^2 $. The previous parameters
give a probability $\sim 1.6\%$ in this case. Despite the fact that the
admixture with $| \psi_{-,+}\rangle$ can also be systematically reduced with 
$|\varepsilon_A|$, this represents a more serious limitation to realize $|-
\rangle \simeq | \psi_{+,-}\rangle$ with high accuracy. 

\section{Derivation of the effective Hamiltonian at the anticrossing point}\label{appendix2}

We present here the derivation of $H_{\rm eff}$ in Eq.~(\ref{H_eff}). 
By using the local spin basis, the Zeeman Hamiltonian has a simple diagonal
form and can be separated as follows: 
\begin{eqnarray}
H_{Z} &=&-|g|\mu _{B}\sum_{l=1,2}b_{l}\tilde{S}_{l,z}  \notag \\
&=&-|g|\mu _{B}b\sum_{l}\tilde{S}_{l,z}-|g|\mu _{B}\frac{\Delta b}{2}(\tilde{%
S}_{1,z}-\tilde{S}_{2,z}),  \label{HZ_rotated}
\end{eqnarray}%
where $\tilde{\mathbf{S}}_{l,z}=\frac{1}{2}(\tilde{d}_{l+}^{\dag
}\tilde{d}_{l+}-\tilde{d}_{l-}^{\dag }\tilde{d}_{l-})$ is the spin operator for the $l$-th dot along
the local field direction. In the second line we have separated the
homogeneous part, proportional to $b$, from the smaller contribution
proportional to $\Delta b$. Following this partition, we write 
$H_{Z}=H_{Z0}+\delta H_{Z}$.

Similarly we define  $\delta H_{T}$ from the tunneling Hamiltonian. Applying the same spin rotation, we write: 
\begin{eqnarray}
H_{T} &=&t\cos (\theta /2)\sum_{\mu =\pm }\left( \tilde{d}_{1\mu }^{\dag }\tilde{d}_{2\mu
}+\tilde{d}_{2\mu }^{\dag }\tilde{d}_{1\mu }\right)  \nonumber \\
&&-t\sin (\theta /2)\sum_{\mu =\pm }\mu \left( \tilde{d}_{1\mu }^{\dag }\tilde{d}_{2\bar{\mu}%
}+\tilde{d}_{2\bar{\mu}}^{\dag }\tilde{d}_{1\mu }\right) ,  \label{HT_rotated}
\end{eqnarray}%
where $\bar{\mu}=-\mu $. The physical meaning of this expression is rather
obvious: electrons maintain the original spin direction upon tunneling but,
due to the different quantization directions on $l=1,2$, the spin appears to
have rotated when expressed through the local spinor basis. Therefore, 
introducing the local quantization axes generates in the
second line of Eq.~(\ref{HT_rotated}) a spin-flip tunneling term analogous to  
the one induced by spin-orbit interaction.\cite{Stepanenko2012} 
For $\theta \ll 1$ we can define the second line of Eq.~
(\ref{HT_rotated}) as a perturbation $\delta H_{T}$ and write $%
H_{T}=H_{T0}+\delta H_{T}$.

The unperturbed Hamiltonian $H_{0}\equiv H_C+ H_{T0}+ H_{Z0}$ is
formally equivalent to the familiar problem of a double dot with
uniform magnetic field of strength $b$ and a modified spin-preserving
tunneling amplitude $t\cos( \theta /2)$. The solution of that problem is well-known in terms of
singlet and triplet states. In particular, the ``triplet'' eigenstates read: 
\begin{eqnarray}
&&|\tilde{T}_{\pm }\rangle =|\tilde{\psi}_{\pm ,\pm }\rangle , \\
&&|\tilde{T}_{0}\rangle =\frac{1}{\sqrt{2}}\left( |\tilde{\psi}_{+,-}\rangle
+|\tilde{\psi}_{-,+}\rangle \right) ,
\end{eqnarray}%
with unperturbed eigenvalues $\mp |g|\mu _{B}b$ and $0$, respectively. The (1,1) ``singlet'' is: 
\begin{equation}
|\tilde{S}(1,1)\rangle =\frac{1}{\sqrt{2}}\left( |\tilde{\psi}_{+,-}\rangle
-|\tilde{\psi}_{-,+}\rangle \right) .
\end{equation}%
Notice that these states differ form the standard singlet/triplets since the
spin quantization axes are different on the two sites $l=1,2$.
On the other hand, $|\tilde{S}(0,2)\rangle $ is the usual singlet state since it involves two
electrons on the same dot. Diagonalization of $H_0$ in the $|\tilde{S}(1,1)\rangle ,|%
\tilde{S}(0,2)\rangle $ subspace yields the eigenstates: 
\begin{equation}
|\tilde{S}_{\pm }\rangle =\sqrt{\frac{\Delta \pm \varepsilon }{2\Delta }}|%
\tilde{S}(1,1)\rangle \pm \sqrt{\frac{\Delta \mp \varepsilon }{2\Delta }}|%
\tilde{S}(0,2)\rangle ,
\label{Stilde_pm}
\end{equation}%
with energy $(-\varepsilon \pm \Delta )/2$, where $\Delta$ is given in Eq.~(\ref{Delta}) of the main text.
For a large region of detunings, these unperturbed eigenstates are a good approximation of
the exact eigenstates. However, the effect of $\delta H_{Z},\delta H_{T}$
becomes important at detunings $\varepsilon _{A,B}$, which
are particularly relevant for our single-spin manipulation scheme.
Around $\varepsilon_B$, it is appropriate to restrict ourselves to the $|\tilde{T}_{+ }\rangle$, $|\tilde{S}_{- }\rangle$ subspace, which gives Eq.~(\ref{H_eff}) of the main text. In particular, the off-diagonal terms in $H_{\rm eff}$ are due to $\delta H_{T}$. 

\bibliography{bibfile}

\begin{thebibliography}{40}%
\makeatletter
\providecommand \@ifxundefined [1]{%
 \@ifx{#1\undefined}
}%
\providecommand \@ifnum [1]{%
 \ifnum #1\expandafter \@firstoftwo
 \else \expandafter \@secondoftwo
 \fi
}%
\providecommand \@ifx [1]{%
 \ifx #1\expandafter \@firstoftwo
 \else \expandafter \@secondoftwo
 \fi
}%
\providecommand \natexlab [1]{#1}%
\providecommand \enquote  [1]{``#1''}%
\providecommand \bibnamefont  [1]{#1}%
\providecommand \bibfnamefont [1]{#1}%
\providecommand \citenamefont [1]{#1}%
\providecommand \href@noop [0]{\@secondoftwo}%
\providecommand \href [0]{\begingroup \@sanitize@url \@href}%
\providecommand \@href[1]{\@@startlink{#1}\@@href}%
\providecommand \@@href[1]{\endgroup#1\@@endlink}%
\providecommand \@sanitize@url [0]{\catcode `\\12\catcode `\$12\catcode
  `\&12\catcode `\#12\catcode `\^12\catcode `\_12\catcode `\%12\relax}%
\providecommand \@@startlink[1]{}%
\providecommand \@@endlink[0]{}%
\providecommand \url  [0]{\begingroup\@sanitize@url \@url }%
\providecommand \@url [1]{\endgroup\@href {#1}{\urlprefix }}%
\providecommand \urlprefix  [0]{URL }%
\providecommand \Eprint [0]{\href }%
\providecommand \doibase [0]{http://dx.doi.org/}%
\providecommand \selectlanguage [0]{\@gobble}%
\providecommand \bibinfo  [0]{\@secondoftwo}%
\providecommand \bibfield  [0]{\@secondoftwo}%
\providecommand \translation [1]{[#1]}%
\providecommand \BibitemOpen [0]{}%
\providecommand \bibitemStop [0]{}%
\providecommand \bibitemNoStop [0]{.\EOS\space}%
\providecommand \EOS [0]{\spacefactor3000\relax}%
\providecommand \BibitemShut  [1]{\csname bibitem#1\endcsname}%
\let\auto@bib@innerbib\@empty
%</preamble>
\bibitem [{\citenamefont {Loss}\ and\ \citenamefont
  {DiVincenzo}(1998)}]{Loss1998}%
  \BibitemOpen
  \bibfield  {author} {\bibinfo {author} {\bibfnamefont {D.}~\bibnamefont
  {Loss}}\ and\ \bibinfo {author} {\bibfnamefont {D.~P.}\ \bibnamefont
  {DiVincenzo}},\ }\href@noop {} {\bibfield  {journal} {\bibinfo  {journal}
  {Phys. Rev. A}\ }\textbf {\bibinfo {volume} {57}},\ \bibinfo {pages} {120}
  (\bibinfo {year} {1998)})}\BibitemShut {NoStop}%
\bibitem [{\citenamefont {\.Zak}\ \emph {et~al.}(2010)\citenamefont {\.Zak},
  \citenamefont {R\"othlisberger}, \citenamefont {Chesi},\ and\ \citenamefont
  {Loss}}]{Zak2010}%
  \BibitemOpen
  \bibfield  {author} {\bibinfo {author} {\bibfnamefont {R.~A.}\ \bibnamefont
  {\.Zak}}, \bibinfo {author} {\bibfnamefont {B.}~\bibnamefont
  {R\"othlisberger}}, \bibinfo {author} {\bibfnamefont {S.}~\bibnamefont
  {Chesi}}, \ and\ \bibinfo {author} {\bibfnamefont {D.}~\bibnamefont {Loss}},\
  }\href@noop {} {\bibfield  {journal} {\bibinfo  {journal} {Riv. Nuovo
  Cimento}\ }\textbf {\bibinfo {volume} {33}},\ \bibinfo {pages} {345}
  (\bibinfo {year} {2010})}\BibitemShut {NoStop}%
\bibitem [{\citenamefont {Kloeffel}\ and\ \citenamefont
  {Loss}(2013)}]{Kloeffel2013}%
  \BibitemOpen
  \bibfield  {author} {\bibinfo {author} {\bibfnamefont {C.}~\bibnamefont
  {Kloeffel}}\ and\ \bibinfo {author} {\bibfnamefont {D.}~\bibnamefont
  {Loss}},\ }\href@noop {} {\bibfield  {journal} {\bibinfo  {journal} {Annu.
  Rev. Condens. Matter Phys.}\ }\textbf {\bibinfo {volume} {4}},\ \bibinfo
  {pages} {51} (\bibinfo {year} {2013})}\BibitemShut {NoStop}%
\bibitem [{\citenamefont {Petta}\ \emph {et~al.}(2005)\citenamefont {Petta},
  \citenamefont {Johnson}, \citenamefont {Taylor}, \citenamefont {Laird},
  \citenamefont {Yacoby}, \citenamefont {Lukin}, \citenamefont {Marcus},
  \citenamefont {Hanson},\ and\ \citenamefont {Gossard}}]{Petta2005}%
  \BibitemOpen
  \bibfield  {author} {\bibinfo {author} {\bibfnamefont {J.~R.}\ \bibnamefont
  {Petta}}, \bibinfo {author} {\bibfnamefont {A.~C.}\ \bibnamefont {Johnson}},
  \bibinfo {author} {\bibfnamefont {J.~M.}\ \bibnamefont {Taylor}}, \bibinfo
  {author} {\bibfnamefont {E.~A.}\ \bibnamefont {Laird}}, \bibinfo {author}
  {\bibfnamefont {A.}~\bibnamefont {Yacoby}}, \bibinfo {author} {\bibfnamefont
  {M.~D.}\ \bibnamefont {Lukin}}, \bibinfo {author} {\bibfnamefont {C.~M.}\
  \bibnamefont {Marcus}}, \bibinfo {author} {\bibfnamefont {M.~P.}\
  \bibnamefont {Hanson}}, \ and\ \bibinfo {author} {\bibfnamefont {A.~C.}\
  \bibnamefont {Gossard}},\ }\href {\doibase 10.1126/science.1116955}
  {\bibfield  {journal} {\bibinfo  {journal} {Science}\ }\textbf {\bibinfo
  {volume} {309}},\ \bibinfo {pages} {2180} (\bibinfo {year}
  {2005})}\BibitemShut {NoStop}%
\bibitem [{\citenamefont {Foletti}\ \emph {et~al.}(2009)\citenamefont
  {Foletti}, \citenamefont {Bluhm}, \citenamefont {Mahalu}, \citenamefont
  {Umansky},\ and\ \citenamefont {Yacoby}}]{Foletti2009}%
  \BibitemOpen
  \bibfield  {author} {\bibinfo {author} {\bibfnamefont {S.}~\bibnamefont
  {Foletti}}, \bibinfo {author} {\bibfnamefont {H.}~\bibnamefont {Bluhm}},
  \bibinfo {author} {\bibfnamefont {D.}~\bibnamefont {Mahalu}}, \bibinfo
  {author} {\bibfnamefont {V.}~\bibnamefont {Umansky}}, \ and\ \bibinfo
  {author} {\bibfnamefont {A.}~\bibnamefont {Yacoby}},\ }\href@noop {}
  {\bibfield  {journal} {\bibinfo  {journal} {Nat. Phys.}\ }\textbf {\bibinfo
  {volume} {5}},\ \bibinfo {pages} {903} (\bibinfo {year} {2009})}\BibitemShut
  {NoStop}%
\bibitem [{\citenamefont {Bluhm}\ \emph {et~al.}(2011)\citenamefont {Bluhm},
  \citenamefont {Foletti}, \citenamefont {Neder}, \citenamefont {Rudner},
  \citenamefont {Mahalu}, \citenamefont {Umansky},\ and\ \citenamefont
  {Yacoby}}]{Bluhm2011}%
  \BibitemOpen
  \bibfield  {author} {\bibinfo {author} {\bibfnamefont {H.}~\bibnamefont
  {Bluhm}}, \bibinfo {author} {\bibfnamefont {S.}~\bibnamefont {Foletti}},
  \bibinfo {author} {\bibfnamefont {I.}~\bibnamefont {Neder}}, \bibinfo
  {author} {\bibfnamefont {M.}~\bibnamefont {Rudner}}, \bibinfo {author}
  {\bibfnamefont {D.}~\bibnamefont {Mahalu}}, \bibinfo {author} {\bibfnamefont
  {V.}~\bibnamefont {Umansky}}, \ and\ \bibinfo {author} {\bibfnamefont
  {A.}~\bibnamefont {Yacoby}},\ }\href@noop {} {\bibfield  {journal} {\bibinfo
  {journal} {Nat. Phys.}\ }\textbf {\bibinfo {volume} {7}},\ \bibinfo {pages}
  {109} (\bibinfo {year} {2011})}\BibitemShut {NoStop}%
\bibitem [{\citenamefont {Taylor}\ \emph {et~al.}(2005)\citenamefont {Taylor},
  \citenamefont {Engel}, \citenamefont {Dur}, \citenamefont {Yacoby},
  \citenamefont {Marcus}, \citenamefont {Zoller},\ and\ \citenamefont
  {Lukin}}]{Taylor2005}%
  \BibitemOpen
  \bibfield  {author} {\bibinfo {author} {\bibfnamefont {J.~M.}\ \bibnamefont
  {Taylor}}, \bibinfo {author} {\bibfnamefont {H.-A.}\ \bibnamefont {Engel}},
  \bibinfo {author} {\bibfnamefont {W.}~\bibnamefont {Dur}}, \bibinfo {author}
  {\bibfnamefont {A.}~\bibnamefont {Yacoby}}, \bibinfo {author} {\bibfnamefont
  {C.~M.}\ \bibnamefont {Marcus}}, \bibinfo {author} {\bibfnamefont
  {P.}~\bibnamefont {Zoller}}, \ and\ \bibinfo {author} {\bibfnamefont {M.~D.}\
  \bibnamefont {Lukin}},\ }\href@noop {} {\bibfield  {journal} {\bibinfo
  {journal} {Nat. Phys.}\ }\textbf {\bibinfo {volume} {1}},\ \bibinfo {pages}
  {177} (\bibinfo {year} {2005})}\BibitemShut {NoStop}%
\bibitem [{\citenamefont {Klinovaja}\ \emph {et~al.}(2012)\citenamefont
  {Klinovaja}, \citenamefont {Stepanenko}, \citenamefont {Halperin},\ and\
  \citenamefont {Loss}}]{Klinovaja2012}%
  \BibitemOpen
  \bibfield  {author} {\bibinfo {author} {\bibfnamefont {J.}~\bibnamefont
  {Klinovaja}}, \bibinfo {author} {\bibfnamefont {D.}~\bibnamefont
  {Stepanenko}}, \bibinfo {author} {\bibfnamefont {B.~I.}\ \bibnamefont
  {Halperin}}, \ and\ \bibinfo {author} {\bibfnamefont {D.}~\bibnamefont
  {Loss}},\ }\href {\doibase 10.1103/PhysRevB.86.085423} {\bibfield  {journal}
  {\bibinfo  {journal} {Phys. Rev. B}\ }\textbf {\bibinfo {volume} {86}},\
  \bibinfo {pages} {085423} (\bibinfo {year} {2012})}\BibitemShut {NoStop}%
\bibitem [{\citenamefont {Shulman}\ \emph {et~al.}(2012)\citenamefont
  {Shulman}, \citenamefont {Dial}, \citenamefont {Harvey}, \citenamefont
  {Bluhm}, \citenamefont {Umansky},\ and\ \citenamefont
  {Yacoby}}]{Shulman2012}%
  \BibitemOpen
  \bibfield  {author} {\bibinfo {author} {\bibfnamefont {M.~D.}\ \bibnamefont
  {Shulman}}, \bibinfo {author} {\bibfnamefont {O.~E.}\ \bibnamefont {Dial}},
  \bibinfo {author} {\bibfnamefont {S.~P.}\ \bibnamefont {Harvey}}, \bibinfo
  {author} {\bibfnamefont {H.}~\bibnamefont {Bluhm}}, \bibinfo {author}
  {\bibfnamefont {V.}~\bibnamefont {Umansky}}, \ and\ \bibinfo {author}
  {\bibfnamefont {A.}~\bibnamefont {Yacoby}},\ }\href@noop {} {\bibfield
  {journal} {\bibinfo  {journal} {Science}\ }\textbf {\bibinfo {volume}
  {336}},\ \bibinfo {pages} {202} (\bibinfo {year} {2012})}\BibitemShut
  {NoStop}%
\bibitem [{\citenamefont {Golovach}\ \emph {et~al.}(2006)\citenamefont
  {Golovach}, \citenamefont {Borhani},\ and\ \citenamefont
  {Loss}}]{Golovach2006}%
  \BibitemOpen
  \bibfield  {author} {\bibinfo {author} {\bibfnamefont {V.~N.}\ \bibnamefont
  {Golovach}}, \bibinfo {author} {\bibfnamefont {M.}~\bibnamefont {Borhani}}, \
  and\ \bibinfo {author} {\bibfnamefont {D.}~\bibnamefont {Loss}},\ }\href@noop
  {} {\bibfield  {journal} {\bibinfo  {journal} {Phys. Rev. B}\ }\textbf
  {\bibinfo {volume} {74}},\ \bibinfo {pages} {165319} (\bibinfo {year}
  {2006})}\BibitemShut {NoStop}%
\bibitem [{\citenamefont {Nowack}\ \emph {et~al.}(2007)\citenamefont {Nowack},
  \citenamefont {Koppens}, \citenamefont {Nazarov},\ and\ \citenamefont
  {Vandersypen}}]{Nowack2007}%
  \BibitemOpen
  \bibfield  {author} {\bibinfo {author} {\bibfnamefont {K.~C.}\ \bibnamefont
  {Nowack}}, \bibinfo {author} {\bibfnamefont {F.~H.~L.}\ \bibnamefont
  {Koppens}}, \bibinfo {author} {\bibfnamefont {Y.~V.}\ \bibnamefont
  {Nazarov}}, \ and\ \bibinfo {author} {\bibfnamefont {L.~M.~K.}\ \bibnamefont
  {Vandersypen}},\ }\href {\doibase 10.1126/science.1148092} {\bibfield
  {journal} {\bibinfo  {journal} {Science}\ }\textbf {\bibinfo {volume}
  {318}},\ \bibinfo {pages} {1430} (\bibinfo {year} {2007})}\BibitemShut
  {NoStop}%
\bibitem [{\citenamefont {Nadj-Perge}\ \emph {et~al.}(2010)\citenamefont
  {Nadj-Perge}, \citenamefont {Frolov}, \citenamefont {Bakkers},\ and\
  \citenamefont {Kouwenhoven}}]{Nadj-Perge2010}%
  \BibitemOpen
  \bibfield  {author} {\bibinfo {author} {\bibfnamefont {S.}~\bibnamefont
  {Nadj-Perge}}, \bibinfo {author} {\bibfnamefont {S.~M.}\ \bibnamefont
  {Frolov}}, \bibinfo {author} {\bibfnamefont {E.~P. A.~M.}\ \bibnamefont
  {Bakkers}}, \ and\ \bibinfo {author} {\bibfnamefont {L.~P.}\ \bibnamefont
  {Kouwenhoven}},\ }\href@noop {} {\bibfield  {journal} {\bibinfo  {journal}
  {Nature}\ }\textbf {\bibinfo {volume} {468}},\ \bibinfo {pages} {1084}
  (\bibinfo {year} {2010})}\BibitemShut {NoStop}%
\bibitem [{\citenamefont {Tokura}\ \emph {et~al.}(2006)\citenamefont {Tokura},
  \citenamefont {van~der Wiel}, \citenamefont {Obata},\ and\ \citenamefont
  {Tarucha}}]{Tokura2006}%
  \BibitemOpen
  \bibfield  {author} {\bibinfo {author} {\bibfnamefont {Y.}~\bibnamefont
  {Tokura}}, \bibinfo {author} {\bibfnamefont {W.~G.}\ \bibnamefont {van~der
  Wiel}}, \bibinfo {author} {\bibfnamefont {T.}~\bibnamefont {Obata}}, \ and\
  \bibinfo {author} {\bibfnamefont {S.}~\bibnamefont {Tarucha}},\ }\href
  {\doibase 10.1103/PhysRevLett.96.047202} {\bibfield  {journal} {\bibinfo
  {journal} {Phys. Rev. Lett.}\ }\textbf {\bibinfo {volume} {96}},\ \bibinfo
  {pages} {047202} (\bibinfo {year} {2006})}\BibitemShut {NoStop}%
\bibitem [{\citenamefont {Pioro-Ladri\`ere}\ \emph {et~al.}(2008)\citenamefont
  {Pioro-Ladri\`ere}, \citenamefont {Obata}, \citenamefont {Tokura},
  \citenamefont {Shin}, \citenamefont {Kubo}, \citenamefont {Yoshida},
  \citenamefont {Taniyama},\ and\ \citenamefont {Tarucha}}]{Pioro2008}%
  \BibitemOpen
  \bibfield  {author} {\bibinfo {author} {\bibfnamefont {M.}~\bibnamefont
  {Pioro-Ladri\`ere}}, \bibinfo {author} {\bibfnamefont {T.}~\bibnamefont
  {Obata}}, \bibinfo {author} {\bibfnamefont {Y.}~\bibnamefont {Tokura}},
  \bibinfo {author} {\bibfnamefont {Y.-S.}\ \bibnamefont {Shin}}, \bibinfo
  {author} {\bibfnamefont {T.}~\bibnamefont {Kubo}}, \bibinfo {author}
  {\bibfnamefont {K.}~\bibnamefont {Yoshida}}, \bibinfo {author} {\bibfnamefont
  {T.}~\bibnamefont {Taniyama}}, \ and\ \bibinfo {author} {\bibfnamefont
  {S.}~\bibnamefont {Tarucha}},\ }\href@noop {} {\bibfield  {journal} {\bibinfo
   {journal} {Nature Phys.}\ }\textbf {\bibinfo {volume} {4}},\ \bibinfo
  {pages} {776} (\bibinfo {year} {2008})}\BibitemShut {NoStop}%
\bibitem [{\citenamefont {Obata}\ \emph {et~al.}(2010)\citenamefont {Obata},
  \citenamefont {Pioro-Ladri\`ere}, \citenamefont {Tokura}, \citenamefont
  {Shin}, \citenamefont {Kubo}, \citenamefont {Yoshida}, \citenamefont
  {Taniyama},\ and\ \citenamefont {Tarucha}}]{Obata2010}%
  \BibitemOpen
  \bibfield  {author} {\bibinfo {author} {\bibfnamefont {T.}~\bibnamefont
  {Obata}}, \bibinfo {author} {\bibfnamefont {M.}~\bibnamefont
  {Pioro-Ladri\`ere}}, \bibinfo {author} {\bibfnamefont {Y.}~\bibnamefont
  {Tokura}}, \bibinfo {author} {\bibfnamefont {Y.-S.}\ \bibnamefont {Shin}},
  \bibinfo {author} {\bibfnamefont {T.}~\bibnamefont {Kubo}}, \bibinfo {author}
  {\bibfnamefont {K.}~\bibnamefont {Yoshida}}, \bibinfo {author} {\bibfnamefont
  {T.}~\bibnamefont {Taniyama}}, \ and\ \bibinfo {author} {\bibfnamefont
  {S.}~\bibnamefont {Tarucha}},\ }\href@noop {} {\bibfield  {journal} {\bibinfo
   {journal} {Phys. Rev. B}\ }\textbf {\bibinfo {volume} {81}},\ \bibinfo
  {pages} {085317} (\bibinfo {year} {2010})}\BibitemShut {NoStop}%
\bibitem [{\citenamefont {Brunner}\ \emph {et~al.}(2011)\citenamefont
  {Brunner}, \citenamefont {Shin}, \citenamefont {Obata}, \citenamefont
  {Pioro-Ladri\`ere}, \citenamefont {Kubo}, \citenamefont {Yoshida},
  \citenamefont {Taniyama}, \citenamefont {Tokura},\ and\ \citenamefont
  {Tarucha}}]{Brunner2011}%
  \BibitemOpen
  \bibfield  {author} {\bibinfo {author} {\bibfnamefont {R.}~\bibnamefont
  {Brunner}}, \bibinfo {author} {\bibfnamefont {Y.-S.}\ \bibnamefont {Shin}},
  \bibinfo {author} {\bibfnamefont {T.}~\bibnamefont {Obata}}, \bibinfo
  {author} {\bibfnamefont {M.}~\bibnamefont {Pioro-Ladri\`ere}}, \bibinfo
  {author} {\bibfnamefont {T.}~\bibnamefont {Kubo}}, \bibinfo {author}
  {\bibfnamefont {K.}~\bibnamefont {Yoshida}}, \bibinfo {author} {\bibfnamefont
  {T.}~\bibnamefont {Taniyama}}, \bibinfo {author} {\bibfnamefont
  {Y.}~\bibnamefont {Tokura}}, \ and\ \bibinfo {author} {\bibfnamefont
  {S.}~\bibnamefont {Tarucha}},\ }\href@noop {} {\bibfield  {journal} {\bibinfo
   {journal} {Phys. Rev. Lett.}\ }\textbf {\bibinfo {volume} {107}},\ \bibinfo
  {pages} {146801} (\bibinfo {year} {2011})}\BibitemShut {NoStop}%
\bibitem [{\citenamefont {Yoneda}\ \emph {et~al.}(2014)\citenamefont {Yoneda},
  \citenamefont {Otsuka}, \citenamefont {Nakajima}, \citenamefont {Takakura},
  \citenamefont {Obata}, \citenamefont {Pioro-Ladri\`ere}, \citenamefont {Lu},
  \citenamefont {Palmstr{\o}m}, \citenamefont {Gossard},\ and\ \citenamefont
  {Tarucha}}]{Tarucha_micromagnet}%
  \BibitemOpen
  \bibfield  {author} {\bibinfo {author} {\bibfnamefont {J.}~\bibnamefont
  {Yoneda}}, \bibinfo {author} {\bibfnamefont {T.}~\bibnamefont {Otsuka}},
  \bibinfo {author} {\bibfnamefont {T.}~\bibnamefont {Nakajima}}, \bibinfo
  {author} {\bibfnamefont {T.}~\bibnamefont {Takakura}}, \bibinfo {author}
  {\bibfnamefont {T.}~\bibnamefont {Obata}}, \bibinfo {author} {\bibfnamefont
  {M.}~\bibnamefont {Pioro-Ladri\`ere}}, \bibinfo {author} {\bibfnamefont
  {H.}~\bibnamefont {Lu}}, \bibinfo {author} {\bibfnamefont {C.}~\bibnamefont
  {Palmstr{\o}m}}, \bibinfo {author} {\bibfnamefont {A.~C.}\ \bibnamefont
  {Gossard}}, \ and\ \bibinfo {author} {\bibfnamefont {S.}~\bibnamefont
  {Tarucha}},\ }\href@noop {} {\bibfield  {journal} {\bibinfo  {journal}
  {arXiv:1411.6738}\ } (\bibinfo {year} {2014})}\BibitemShut {NoStop}%
\bibitem [{\citenamefont {Coish}\ and\ \citenamefont {Loss}(2007)}]{Coish2007}%
  \BibitemOpen
  \bibfield  {author} {\bibinfo {author} {\bibfnamefont {W.~A.}\ \bibnamefont
  {Coish}}\ and\ \bibinfo {author} {\bibfnamefont {D.}~\bibnamefont {Loss}},\
  }\href {\doibase 10.1103/PhysRevB.75.161302} {\bibfield  {journal} {\bibinfo
  {journal} {Phys. Rev. B}\ }\textbf {\bibinfo {volume} {75}},\ \bibinfo
  {pages} {161302} (\bibinfo {year} {2007})}\BibitemShut {NoStop}%
\bibitem [{\citenamefont {Trifunovic}\ \emph {et~al.}(2012)\citenamefont
  {Trifunovic}, \citenamefont {Dial}, \citenamefont {Trif}, \citenamefont
  {Wootton}, \citenamefont {Abebe}, \citenamefont {Yacoby},\ and\ \citenamefont
  {Loss}}]{Trifunovic2012}%
  \BibitemOpen
  \bibfield  {author} {\bibinfo {author} {\bibfnamefont {L.}~\bibnamefont
  {Trifunovic}}, \bibinfo {author} {\bibfnamefont {O.}~\bibnamefont {Dial}},
  \bibinfo {author} {\bibfnamefont {M.}~\bibnamefont {Trif}}, \bibinfo {author}
  {\bibfnamefont {J.~R.}\ \bibnamefont {Wootton}}, \bibinfo {author}
  {\bibfnamefont {R.}~\bibnamefont {Abebe}}, \bibinfo {author} {\bibfnamefont
  {A.}~\bibnamefont {Yacoby}}, \ and\ \bibinfo {author} {\bibfnamefont
  {D.}~\bibnamefont {Loss}},\ }\href {\doibase 10.1103/PhysRevX.2.011006}
  {\bibfield  {journal} {\bibinfo  {journal} {Phys. Rev. X}\ }\textbf {\bibinfo
  {volume} {2}},\ \bibinfo {pages} {011006} (\bibinfo {year}
  {2012})}\BibitemShut {NoStop}%
\bibitem [{\citenamefont {Trifunovic}\ \emph {et~al.}(2013)\citenamefont
  {Trifunovic}, \citenamefont {Pedrocchi},\ and\ \citenamefont
  {Loss}}]{Trifunovic2013}%
  \BibitemOpen
  \bibfield  {author} {\bibinfo {author} {\bibfnamefont {L.}~\bibnamefont
  {Trifunovic}}, \bibinfo {author} {\bibfnamefont {F.~L.}\ \bibnamefont
  {Pedrocchi}}, \ and\ \bibinfo {author} {\bibfnamefont {D.}~\bibnamefont
  {Loss}},\ }\href {\doibase 10.1103/PhysRevX.3.041023} {\bibfield  {journal}
  {\bibinfo  {journal} {Phys. Rev. X}\ }\textbf {\bibinfo {volume} {3}},\
  \bibinfo {pages} {041023} (\bibinfo {year} {2013})}\BibitemShut {NoStop}%
\bibitem [{\citenamefont {Ribeiro}\ \emph {et~al.}(2010)\citenamefont
  {Ribeiro}, \citenamefont {Petta},\ and\ \citenamefont
  {Burkard}}]{Ribeiro2010}%
  \BibitemOpen
  \bibfield  {author} {\bibinfo {author} {\bibfnamefont {H.}~\bibnamefont
  {Ribeiro}}, \bibinfo {author} {\bibfnamefont {J.~R.}\ \bibnamefont {Petta}},
  \ and\ \bibinfo {author} {\bibfnamefont {G.}~\bibnamefont {Burkard}},\
  }\href@noop {} {\bibfield  {journal} {\bibinfo  {journal} {Phys. Rev. B}\
  }\textbf {\bibinfo {volume} {82}},\ \bibinfo {pages} {115445} (\bibinfo
  {year} {2010})}\BibitemShut {NoStop}%
\bibitem [{\citenamefont {Petta}\ \emph {et~al.}(2010)\citenamefont {Petta},
  \citenamefont {Lu},\ and\ \citenamefont {Gossard}}]{Petta2010}%
  \BibitemOpen
  \bibfield  {author} {\bibinfo {author} {\bibfnamefont {R.}~\bibnamefont
  {Petta}}, \bibinfo {author} {\bibfnamefont {H.}~\bibnamefont {Lu}}, \ and\
  \bibinfo {author} {\bibfnamefont {A.~C.}\ \bibnamefont {Gossard}},\
  }\href@noop {} {\bibfield  {journal} {\bibinfo  {journal} {Science}\ }\textbf
  {\bibinfo {volume} {327}},\ \bibinfo {pages} {669} (\bibinfo {year}
  {2010})}\BibitemShut {NoStop}%
\bibitem [{\citenamefont {Ribeiro}\ \emph
  {et~al.}(2013{\natexlab{a}})\citenamefont {Ribeiro}, \citenamefont {Burkard},
  \citenamefont {Petta}, \citenamefont {Lu},\ and\ \citenamefont
  {Gossard}}]{Ribeiro2013a}%
  \BibitemOpen
  \bibfield  {author} {\bibinfo {author} {\bibfnamefont {H.}~\bibnamefont
  {Ribeiro}}, \bibinfo {author} {\bibfnamefont {G.}~\bibnamefont {Burkard}},
  \bibinfo {author} {\bibfnamefont {J.~R.}\ \bibnamefont {Petta}}, \bibinfo
  {author} {\bibfnamefont {H.}~\bibnamefont {Lu}}, \ and\ \bibinfo {author}
  {\bibfnamefont {A.~C.}\ \bibnamefont {Gossard}},\ }\href {\doibase
  10.1103/PhysRevLett.110.086804} {\bibfield  {journal} {\bibinfo  {journal}
  {Phys. Rev. Lett.}\ }\textbf {\bibinfo {volume} {110}},\ \bibinfo {pages}
  {086804} (\bibinfo {year} {2013}{\natexlab{a}})}\BibitemShut {NoStop}%
\bibitem [{\citenamefont {Ribeiro}\ \emph
  {et~al.}(2013{\natexlab{b}})\citenamefont {Ribeiro}, \citenamefont {Petta},\
  and\ \citenamefont {Burkard}}]{Ribeiro2013b}%
  \BibitemOpen
  \bibfield  {author} {\bibinfo {author} {\bibfnamefont {H.}~\bibnamefont
  {Ribeiro}}, \bibinfo {author} {\bibfnamefont {J.~R.}\ \bibnamefont {Petta}},
  \ and\ \bibinfo {author} {\bibfnamefont {G.}~\bibnamefont {Burkard}},\ }\href
  {\doibase 10.1103/PhysRevB.87.235318} {\bibfield  {journal} {\bibinfo
  {journal} {Phys. Rev. B}\ }\textbf {\bibinfo {volume} {87}},\ \bibinfo
  {pages} {235318} (\bibinfo {year} {2013}{\natexlab{b}})}\BibitemShut
  {NoStop}%
\bibitem [{\citenamefont {Wu}\ \emph {et~al.}(2014)\citenamefont {Wu},
  \citenamefont {Ward}, \citenamefont {Prance}, \citenamefont {Kim},
  \citenamefont {Gamble}, \citenamefont {Mohr}, \citenamefont {Shi},
  \citenamefont {Savage}, \citenamefont {Lagally}, \citenamefont {Friesen},
  \citenamefont {Coppersmith},\ and\ \citenamefont {Eriksson}}]{Wu2014}%
  \BibitemOpen
  \bibfield  {author} {\bibinfo {author} {\bibfnamefont {X.}~\bibnamefont
  {Wu}}, \bibinfo {author} {\bibfnamefont {D.~R.}\ \bibnamefont {Ward}},
  \bibinfo {author} {\bibfnamefont {J.~R.}\ \bibnamefont {Prance}}, \bibinfo
  {author} {\bibfnamefont {D.}~\bibnamefont {Kim}}, \bibinfo {author}
  {\bibfnamefont {J.~K.}\ \bibnamefont {Gamble}}, \bibinfo {author}
  {\bibfnamefont {R.}~\bibnamefont {Mohr}}, \bibinfo {author} {\bibfnamefont
  {Z.}~\bibnamefont {Shi}}, \bibinfo {author} {\bibfnamefont {D.~E.}\
  \bibnamefont {Savage}}, \bibinfo {author} {\bibfnamefont {M.~G.}\
  \bibnamefont {Lagally}}, \bibinfo {author} {\bibfnamefont {M.}~\bibnamefont
  {Friesen}}, \bibinfo {author} {\bibfnamefont {S.~N.}\ \bibnamefont
  {Coppersmith}}, \ and\ \bibinfo {author} {\bibfnamefont {M.~A.}\ \bibnamefont
  {Eriksson}},\ }\href@noop {} {\bibfield  {journal} {\bibinfo  {journal}
  {Proc. Natl. Acad. Sci. USA}\ }\textbf {\bibinfo {volume} {111}},\ \bibinfo
  {pages} {11938} (\bibinfo {year} {2014})}\BibitemShut {NoStop}%
\bibitem [{\citenamefont {Burkard}\ \emph {et~al.}(1999)\citenamefont
  {Burkard}, \citenamefont {Loss},\ and\ \citenamefont
  {DiVincenzo}}]{Burkard1999}%
  \BibitemOpen
  \bibfield  {author} {\bibinfo {author} {\bibfnamefont {G.}~\bibnamefont
  {Burkard}}, \bibinfo {author} {\bibfnamefont {D.}~\bibnamefont {Loss}}, \
  and\ \bibinfo {author} {\bibfnamefont {D.~P.}\ \bibnamefont {DiVincenzo}},\
  }\href@noop {} {\bibfield  {journal} {\bibinfo  {journal} {Phys. Rev. B}\
  }\textbf {\bibinfo {volume} {59}},\ \bibinfo {pages} {2070} (\bibinfo {year}
  {1999})}\BibitemShut {NoStop}%
\bibitem [{\citenamefont {Stepanenko}\ \emph {et~al.}(2012)\citenamefont
  {Stepanenko}, \citenamefont {Rudner}, \citenamefont {Halperin},\ and\
  \citenamefont {Loss}}]{Stepanenko2012}%
  \BibitemOpen
  \bibfield  {author} {\bibinfo {author} {\bibfnamefont {D.}~\bibnamefont
  {Stepanenko}}, \bibinfo {author} {\bibfnamefont {M.}~\bibnamefont {Rudner}},
  \bibinfo {author} {\bibfnamefont {B.~I.}\ \bibnamefont {Halperin}}, \ and\
  \bibinfo {author} {\bibfnamefont {D.}~\bibnamefont {Loss}},\ }\href {\doibase
  10.1103/PhysRevB.85.075416} {\bibfield  {journal} {\bibinfo  {journal} {Phys.
  Rev. B}\ }\textbf {\bibinfo {volume} {85}},\ \bibinfo {pages} {075416}
  (\bibinfo {year} {2012})}\BibitemShut {NoStop}%
\bibitem [{\citenamefont {van~der Wiel}\ \emph {et~al.}(2002)\citenamefont
  {van~der Wiel}, \citenamefont {De~Franceschi}, \citenamefont {Elzerman},
  \citenamefont {Fujisawa}, \citenamefont {Tarucha},\ and\ \citenamefont
  {Kouwenhoven}}]{vdWiel2002}%
  \BibitemOpen
  \bibfield  {author} {\bibinfo {author} {\bibfnamefont {W.~G.}\ \bibnamefont
  {van~der Wiel}}, \bibinfo {author} {\bibfnamefont {S.}~\bibnamefont
  {De~Franceschi}}, \bibinfo {author} {\bibfnamefont {J.~M.}\ \bibnamefont
  {Elzerman}}, \bibinfo {author} {\bibfnamefont {T.}~\bibnamefont {Fujisawa}},
  \bibinfo {author} {\bibfnamefont {S.}~\bibnamefont {Tarucha}}, \ and\
  \bibinfo {author} {\bibfnamefont {L.~P.}\ \bibnamefont {Kouwenhoven}},\
  }\href {\doibase 10.1103/RevModPhys.75.1} {\bibfield  {journal} {\bibinfo
  {journal} {Rev. Mod. Phys.}\ }\textbf {\bibinfo {volume} {75}},\ \bibinfo
  {pages} {1} (\bibinfo {year} {2002})}\BibitemShut {NoStop}%
\bibitem [{\citenamefont {Obata}\ \emph {et~al.}(2012)\citenamefont {Obata},
  \citenamefont {Pioro-Ladri\`ere}, \citenamefont {Tokura},\ and\ \citenamefont
  {Tarucha}}]{Obata2012}%
  \BibitemOpen
  \bibfield  {author} {\bibinfo {author} {\bibfnamefont {T.}~\bibnamefont
  {Obata}}, \bibinfo {author} {\bibfnamefont {M.}~\bibnamefont
  {Pioro-Ladri\`ere}}, \bibinfo {author} {\bibfnamefont {Y.}~\bibnamefont
  {Tokura}}, \ and\ \bibinfo {author} {\bibfnamefont {S.}~\bibnamefont
  {Tarucha}},\ }\href {http://stacks.iop.org/1367-2630/14/i=12/a=123013}
  {\bibfield  {journal} {\bibinfo  {journal} {New J. Phys.}\ }\textbf {\bibinfo
  {volume} {14}},\ \bibinfo {pages} {123013} (\bibinfo {year}
  {2012})}\BibitemShut {NoStop}%
\bibitem [{Rad()}]{Radia}%
  \BibitemOpen
  \href@noop {} {}\bibinfo {note} {RADIA Technical Reference Manual ESRF,
  Grenoble, France.}\BibitemShut {Stop}%
\bibitem [{\citenamefont {Beaudoin}\ and\ \citenamefont
  {Coish}(2013)}]{Beaudoin2013}%
  \BibitemOpen
  \bibfield  {author} {\bibinfo {author} {\bibfnamefont {F.}~\bibnamefont
  {Beaudoin}}\ and\ \bibinfo {author} {\bibfnamefont {W.~A.}\ \bibnamefont
  {Coish}},\ }\href@noop {} {\bibfield  {journal} {\bibinfo  {journal} {Phys.
  Rev. B}\ }\textbf {\bibinfo {volume} {88}},\ \bibinfo {pages} {085320}
  (\bibinfo {year} {2013})}\BibitemShut {NoStop}%
\bibitem [{\citenamefont {Coish}\ and\ \citenamefont {Loss}(2004)}]{Coish2004}%
  \BibitemOpen
  \bibfield  {author} {\bibinfo {author} {\bibfnamefont {W.~A.}\ \bibnamefont
  {Coish}}\ and\ \bibinfo {author} {\bibfnamefont {D.}~\bibnamefont {Loss}},\
  }\href@noop {} {\bibfield  {journal} {\bibinfo  {journal} {Phys. Rev. B}\
  }\textbf {\bibinfo {volume} {70}},\ \bibinfo {pages} {195340} (\bibinfo
  {year} {2004})}\BibitemShut {NoStop}%
\bibitem [{\citenamefont {Dial}\ \emph {et~al.}(2013)\citenamefont {Dial},
  \citenamefont {Shulman}, \citenamefont {Harvey}, \citenamefont {Bluhm},
  \citenamefont {Umansky},\ and\ \citenamefont {Yacoby}}]{Dial2013}%
  \BibitemOpen
  \bibfield  {author} {\bibinfo {author} {\bibfnamefont {O.~E.}\ \bibnamefont
  {Dial}}, \bibinfo {author} {\bibfnamefont {M.~D.}\ \bibnamefont {Shulman}},
  \bibinfo {author} {\bibfnamefont {S.~P.}\ \bibnamefont {Harvey}}, \bibinfo
  {author} {\bibfnamefont {H.}~\bibnamefont {Bluhm}}, \bibinfo {author}
  {\bibfnamefont {V.}~\bibnamefont {Umansky}}, \ and\ \bibinfo {author}
  {\bibfnamefont {A.}~\bibnamefont {Yacoby}},\ }\href@noop {} {\bibfield
  {journal} {\bibinfo  {journal} {Phys. Rev. Lett.}\ }\textbf {\bibinfo
  {volume} {110}},\ \bibinfo {pages} {146804} (\bibinfo {year}
  {2013})}\BibitemShut {NoStop}%
\bibitem [{\citenamefont {Kornich}\ \emph {et~al.}(2014)\citenamefont
  {Kornich}, \citenamefont {Kloeffel},\ and\ \citenamefont
  {Loss}}]{Kornich2014}%
  \BibitemOpen
  \bibfield  {author} {\bibinfo {author} {\bibfnamefont {V.}~\bibnamefont
  {Kornich}}, \bibinfo {author} {\bibfnamefont {C.}~\bibnamefont {Kloeffel}}, \
  and\ \bibinfo {author} {\bibfnamefont {D.}~\bibnamefont {Loss}},\ }\href
  {\doibase 10.1103/PhysRevB.89.085410} {\bibfield  {journal} {\bibinfo
  {journal} {Phys. Rev. B}\ }\textbf {\bibinfo {volume} {89}},\ \bibinfo
  {pages} {085410} (\bibinfo {year} {2014})}\BibitemShut {NoStop}%
\bibitem [{\citenamefont {Koppens}\ \emph {et~al.}(2007)\citenamefont
  {Koppens}, \citenamefont {Klauser}, \citenamefont {Coish}, \citenamefont
  {Nowack}, \citenamefont {Kouwenhoven}, \citenamefont {Loss},\ and\
  \citenamefont {Vandersypen}}]{Koppens2007}%
  \BibitemOpen
  \bibfield  {author} {\bibinfo {author} {\bibfnamefont {F.~H.~L.}\
  \bibnamefont {Koppens}}, \bibinfo {author} {\bibfnamefont {D.}~\bibnamefont
  {Klauser}}, \bibinfo {author} {\bibfnamefont {W.~A.}\ \bibnamefont {Coish}},
  \bibinfo {author} {\bibfnamefont {K.~C.}\ \bibnamefont {Nowack}}, \bibinfo
  {author} {\bibfnamefont {L.~P.}\ \bibnamefont {Kouwenhoven}}, \bibinfo
  {author} {\bibfnamefont {D.}~\bibnamefont {Loss}}, \ and\ \bibinfo {author}
  {\bibfnamefont {L.~M.~K.}\ \bibnamefont {Vandersypen}},\ }\href {\doibase
  10.1103/PhysRevLett.99.106803} {\bibfield  {journal} {\bibinfo  {journal}
  {Phys. Rev. Lett.}\ }\textbf {\bibinfo {volume} {99}},\ \bibinfo {pages}
  {106803} (\bibinfo {year} {2007})}\BibitemShut {NoStop}%
\bibitem [{\citenamefont {Khaetskii}\ and\ \citenamefont
  {Nazarov}(2001)}]{Khaetskii2001}%
  \BibitemOpen
  \bibfield  {author} {\bibinfo {author} {\bibfnamefont {A.~V.}\ \bibnamefont
  {Khaetskii}}\ and\ \bibinfo {author} {\bibfnamefont {Y.~V.}\ \bibnamefont
  {Nazarov}},\ }\href@noop {} {\bibfield  {journal} {\bibinfo  {journal} {Phys.
  Rev. B}\ }\textbf {\bibinfo {volume} {64}},\ \bibinfo {pages} {125316}
  (\bibinfo {year} {2001})}\BibitemShut {NoStop}%
\bibitem [{\citenamefont {Golovach}\ \emph {et~al.}(2004)\citenamefont
  {Golovach}, \citenamefont {Khaetskii},\ and\ \citenamefont
  {Loss}}]{Golovach2004}%
  \BibitemOpen
  \bibfield  {author} {\bibinfo {author} {\bibfnamefont {V.~N.}\ \bibnamefont
  {Golovach}}, \bibinfo {author} {\bibfnamefont {A.}~\bibnamefont {Khaetskii}},
  \ and\ \bibinfo {author} {\bibfnamefont {D.}~\bibnamefont {Loss}},\ }\href
  {\doibase 10.1103/PhysRevLett.93.016601} {\bibfield  {journal} {\bibinfo
  {journal} {Phys. Rev. Lett.}\ }\textbf {\bibinfo {volume} {93}},\ \bibinfo
  {pages} {016601} (\bibinfo {year} {2004})}\BibitemShut {NoStop}%
\bibitem [{\citenamefont {Stano}\ and\ \citenamefont
  {Fabian}(2006)}]{Stano2006}%
  \BibitemOpen
  \bibfield  {author} {\bibinfo {author} {\bibfnamefont {P.}~\bibnamefont
  {Stano}}\ and\ \bibinfo {author} {\bibfnamefont {J.}~\bibnamefont {Fabian}},\
  }\href {\doibase 10.1103/PhysRevLett.96.186602} {\bibfield  {journal}
  {\bibinfo  {journal} {Phys. Rev. Lett.}\ }\textbf {\bibinfo {volume} {96}},\
  \bibinfo {pages} {186602} (\bibinfo {year} {2006})}\BibitemShut {NoStop}%
\bibitem [{\citenamefont {Amasha}\ \emph {et~al.}(2008)\citenamefont {Amasha},
  \citenamefont {MacLean}, \citenamefont {Radu}, \citenamefont {Zumb\"uhl},
  \citenamefont {Kastner}, \citenamefont {Hanson},\ and\ \citenamefont
  {Gossard}}]{Amasha2008}%
  \BibitemOpen
  \bibfield  {author} {\bibinfo {author} {\bibfnamefont {S.}~\bibnamefont
  {Amasha}}, \bibinfo {author} {\bibfnamefont {K.}~\bibnamefont {MacLean}},
  \bibinfo {author} {\bibfnamefont {I.~P.}\ \bibnamefont {Radu}}, \bibinfo
  {author} {\bibfnamefont {D.~M.}\ \bibnamefont {Zumb\"uhl}}, \bibinfo {author}
  {\bibfnamefont {M.~A.}\ \bibnamefont {Kastner}}, \bibinfo {author}
  {\bibfnamefont {M.~P.}\ \bibnamefont {Hanson}}, \ and\ \bibinfo {author}
  {\bibfnamefont {A.~C.}\ \bibnamefont {Gossard}},\ }\href {\doibase
  10.1103/PhysRevLett.100.046803} {\bibfield  {journal} {\bibinfo  {journal}
  {Phys. Rev. Lett.}\ }\textbf {\bibinfo {volume} {100}},\ \bibinfo {pages}
  {046803} (\bibinfo {year} {2008})}\BibitemShut {NoStop}%
\bibitem [{\citenamefont {Scarlino}\ \emph {et~al.}(2014)\citenamefont
  {Scarlino}, \citenamefont {Kawakami}, \citenamefont {Stano}, \citenamefont
  {Shafiei}, \citenamefont {Reichl}, \citenamefont {Wegscheider},\ and\
  \citenamefont {Vandersypen}}]{Scarlino2014}%
  \BibitemOpen
  \bibfield  {author} {\bibinfo {author} {\bibfnamefont {P.}~\bibnamefont
  {Scarlino}}, \bibinfo {author} {\bibfnamefont {E.}~\bibnamefont {Kawakami}},
  \bibinfo {author} {\bibfnamefont {P.}~\bibnamefont {Stano}}, \bibinfo
  {author} {\bibfnamefont {M.}~\bibnamefont {Shafiei}}, \bibinfo {author}
  {\bibfnamefont {C.}~\bibnamefont {Reichl}}, \bibinfo {author} {\bibfnamefont
  {W.}~\bibnamefont {Wegscheider}}, \ and\ \bibinfo {author} {\bibfnamefont
  {L.~M.~K.}\ \bibnamefont {Vandersypen}},\ }\href@noop {} {\bibfield
  {journal} {\bibinfo  {journal} {arXiv:1409.1016}\ } (\bibinfo {year}
  {2014})}\BibitemShut {NoStop}%
\end{thebibliography}%

\end{document}